# Multi-User MISO Interference Channels with Single-User Detection: Optimality of Beamforming and the Achievable Rate Region

Xiaohu Shang, Biao Chen, and H. Vincent Poor

**Abstract**

For a multi-user interference channel with multi-antenna transmitters and single-antenna receivers, by restricting each transmitter to Gaussian input and each receiver to a single-user detector, computing the largest achievable rate region amounts to solving a family of non-convex optimization problems. Recognizing the intrinsic connection between the signal power at the intended receiver and the interference power at the unintended receiver, the original family of non-convex optimization problems is converted into a new family of convex optimization problems. It is shown that, for such interference channels with each receiver implementing single-user detection, transmitter beamforming can achieve all boundary points of the achievable rate region.

## I. Introduction

The interference channel (IC) models a multi-user communication system in which each transmitter communicates to its intended receiver while generating interference to all unintended receivers. Determining the capacity region of an IC remains an open problem except in the case of ICs with strong interference [1], [2]. To date, the best achievable rate region was established by Han and Kobayashi in [1], herein termed the HK region, which combines rate splitting at transmitters, joint decoding at receivers, and time sharing among codebooks. The HK region was simplified by Chong, Motani, Garg,

X. Shang and H. V. Poor are with Princeton University, Department of Electrical Engineering, Princeton, NJ, 08544. Email: xshang@princeton.edu, poor@princeton.edu. B. Chen is with Syracuse University, Department of Electrical Engineering and Computer Science, 335 Link Hall, Syracuse, NY 13244. Email: bichen@syr.edu.

[0]This work was supported in part by the National Science Foundation under Grants CNS-06-25637 and CCF-05-46491.

and El Gamal [3] and several computable subregions were also proposed in [4]–[6]. Etkin, Tse, and Wang [7, Theorem 1] proved that the HK region is within 1-bit of the capacity region of the Gaussian IC. The results in [6], [8] and [9], whose genie-aided approach is largely motivated by [7], established the sum-rate capacity of the two-user Gaussian IC in the low interference regime: when the interference power is below a certain threshold (referred to as noisy interference in [8]), the results assert the *optimality of treating interference as noise at both receivers*, i.e., each receiver should simply implement singe-user detection (SUD). In addition, even if the noisy interference condition is not satisfied, practical constraints often limit the receivers to implementing SUD. For example, the receivers may know only the channels associated with their own intended links. Under such scenarios, treating interference as noise at each receiver is more practical.

There have been several recent studies of multiple-input multiple-output (MIMO) and multiple-input single-output (MISO) ICs [10]–[15]. The MISO IC describes, for example, the downlink communications of co-channel cells where the base stations have multiple antennas and the mobile stations have single antennas. The downlink beamforming problem has been well studied in [16]–[19]. Less well understood is the downlink transmission in the presence of interference, both in terms of its fundamental performance limits (i.e., capacity region), as well as in practically feasible transmission schemes. The assumption of multi-antenna transmitters and single-antenna receivers is motivated by the real world constraints where miniaturization of mobile units limits the number of antennas. In addition, the asymmetry in available resources at base and mobile stations favors systems where transmitters are tasked with heavy processing in exchange for reduced complexity at mobile units. Toward this end, we assume in the present work that each receiver implements SUD, i.e., it treats interference as channel noise. In a preliminary work [12], we showed that beamforming is optimal for the entire SUD rate region for a two-user real MISO IC. This result was used in [20] to characterize the beamforming vectors that achieve the boundary rate points on the SUD rate region. Later, the result in [12] was also used in [15] to derive the noisy-interference sum-rate capacity of the symmetric real MISO IC. In this paper, we generalize the result of [12] to complex multi-user MISO ICs. We note that the proof in [12] is applicable only to two-user real MISO ICs.

There have been various studies concerning throughput optimization in a multi-user system under the assumption that *each receiver treats interference as channel noise* [21]–[26]. However, even for the simple scalar Gaussian IC, computing the largest achievable rate region with single-user detection at each receiver is in general an open problem [27]. Exhaustive search over the transmitter power is typically unavoidable due to the non-convexity of the problem. The difficulty is much more acute for the MISO IC case as one needs to exhaust over all covariance matrices satisfying the power constraints. The complexity increases with the square of the number of transmit antennas, which renders the computation intractable. In this paper we propose an alternative way of deriving the optimal signaling for the SUD rate region for multi-user complex MISO ICs. Our approach is to convert a family of non-convex optimization problems for the original formulation to an equivalent family of convex optimization problems. What is more significant is that, given that each receiver implements SUD, all boundary points of the rate region can be achieved by transmitter beamforming.

The rest of the paper is organized as follows. In Section II, we use the two-user complex MISO IC as an example to explain the basic idea of problem reformulation. We show that beamforming is optimal for the SUD rate region of such channel. The closed-form rate region is also presented. These results are generalized to multi-user complex MISO ICs in Section III. We prove that beamforming is also optimal for $m$-user complex MISO ICs with $m \geq 2$. Based on this result, we use the three-user MISO IC as an example to show how to obtain the SUD rate region for an $m$-user MISO IC. Numerical examples are provided in Section IV. We conclude in Section V.

Before proceeding, we introduce the following notation.

- Bold fact letters, e.g. $\boldsymbol{x}$ and $\mathbf{X}$, denote vectors and matrices respectively.
- $(\cdot)^T$ and $(\cdot)^\dagger$ denote respectively the transpose and the Hermitian (conjugate transpose) of a matrix or a vector. Consequently the Hermitian of a scalar is its conjugate.
- $\mathbf{I}$ is an identity matrix, $\mathbf{0}$ is an all-zero vector or matrix depending on the context, and diag$(\boldsymbol{x})$ is a diagonal matrix with its diagonal entries the same as that of the vector $\boldsymbol{x}$.
- $\mathbf{X} \succeq \mathbf{0}$ means that $\mathbf{X}$ is a positive semi-definite Hermitian matrix.
- tr$(\mathbf{X})$ and rank$(\mathbf{X})$ denote the trace and the rank, respectively, of the matrix $\mathbf{X}$.
- $(\boldsymbol{x})_i$ denotes the $i$th entry of vector $\boldsymbol{x}$, $(\mathbf{X})_{ij}$ denotes the $i$th row and $j$th column entry of matrix $\mathbf{X}$, and $\mathbf{X}_{m\times n}$ means that $\mathbf{X}$ is an $m \times n$ matrix.
- $|x|$ is the absolute value of a scalar $x$, and $\|\boldsymbol{x}\|$ is the norm of a vector $\boldsymbol{x}$, i.e., $\|\boldsymbol{x}\| = \sqrt{\boldsymbol{x}^\dagger \boldsymbol{x}}$.
- $\angle(\boldsymbol{x}, \boldsymbol{y})$ denotes the angle between two real vectors $\boldsymbol{x}$ and $\boldsymbol{y}$, and $\angle(\boldsymbol{x}, \boldsymbol{y}) \in [0, \pi]$. If both $\boldsymbol{x}$ and $\boldsymbol{y}$ are non-zero, then $\angle(\boldsymbol{x}, \boldsymbol{y}) = \cos^{-1} \frac{\boldsymbol{x}^T \boldsymbol{y}}{\|\boldsymbol{x}\| \cdot \|\boldsymbol{y}\|}$. Otherwise we let $\angle(\boldsymbol{x}, \boldsymbol{y}) = \frac{\pi}{2}$ for convenience.
- $E[\cdot]$ denotes expectation.
- sign$(x)$ is the sign of a real scalar $x$, i.e.,

$$\text{sign}(x) = \begin{cases} -1, & \text{if} \quad x < 0, \\ 1, & \text{if} \quad x \geq 0. \end{cases}$$

- $\mathrm{Re}(\cdot)$ denotes the real part of it complex argument.

## II. Two-user MISO IC with single user detector

The two-user Gaussian MISO IC is illustrated in Fig. 1 and the received signals are defined as

$$\begin{aligned} & Y_1 = \boldsymbol{h}_1^\dagger \boldsymbol{x}_1 + \boldsymbol{h}_2^\dagger \boldsymbol{x}_2 + N_1 \\ \text{and} \quad & Y_2 = \boldsymbol{h}_3^\dagger \boldsymbol{x}_1 + \boldsymbol{h}_4^\dagger \boldsymbol{x}_2 + N_2, \end{aligned} \tag{1}$$

where $\boldsymbol{x}_1$ and $\boldsymbol{x}_2$ are transmitted signal vectors of user 1 and user 2 with dimensions $t_1$ and $t_2$, respectively; $Y_1$ and $Y_2$ are two scalar received signals; $N_1$ and $N_2$ are unit variance circularly symmetric complex Gaussian noises; $\boldsymbol{h}_1$ and $\boldsymbol{h}_3$ are $t_1 \times 1$ complex channel vectors; and $\boldsymbol{h}_2$ and $\boldsymbol{h}_4$ are $t_2 \times 1$ complex channel vectors. The power constraints at the transmitters are respectively $\mathrm{tr}(\mathbf{S}_1) \leq P_1$ and $\mathrm{tr}(\mathbf{S}_2) \leq P_2$, where $\mathbf{S}_1 = E\left[\boldsymbol{x}_1 \boldsymbol{x}_1^\dagger\right]$, $\mathbf{S}_2 = E\left[\boldsymbol{x}_2 \boldsymbol{x}_2^\dagger\right]$. We assume that the transmitted signals $\boldsymbol{x}_1$ and $\boldsymbol{x}_2$ are zero-mean Gaussian vectors. Each transmitter knows all the channel vectors. Each receiver knows only the channel vector from its transmitter: receiver 1 (resp. 2) only knows $\boldsymbol{h}_1$ (resp. $\boldsymbol{h}_4$). Each receiver decodes its own signal while treating the interference from the other user as noise. The boundary points of the achievable rate region for this channel are characterized by the following family of optimization problems:

$$\begin{aligned} \max \quad & \mu_1 R_1 + \mu_2 R_2 \\ \text{subject to} \quad & R_1 = \log\left(1 + \frac{\boldsymbol{h}_1^\dagger \mathbf{S}_1 \boldsymbol{h}_1}{1 + \boldsymbol{h}_2^\dagger \mathbf{S}_2 \boldsymbol{h}_2}\right), \\ & R_2 = \log\left(1 + \frac{\boldsymbol{h}_4^\dagger \mathbf{S}_2 \boldsymbol{h}_4}{1 + \boldsymbol{h}_3^\dagger \mathbf{S}_1 \boldsymbol{h}_3}\right), \\ & \mathrm{tr}\left(\mathbf{S}_i\right) \leq P_i, \\ & \mathbf{S}_i \succeq 0, \quad i = 1, 2, \end{aligned} \tag{2}$$

where $0 \leq \mu_1 < \infty$ and $0 \leq \mu_2 < \infty$. We define the SUD rate region of a two-user MISO IC as

$$\bigcup_{\text{all } \mu_1,\mu_2 \geq 0} \left\{ R_1 \text{ and } R_2 \text{ optimal for problem (2)} \right\}. \tag{3}$$

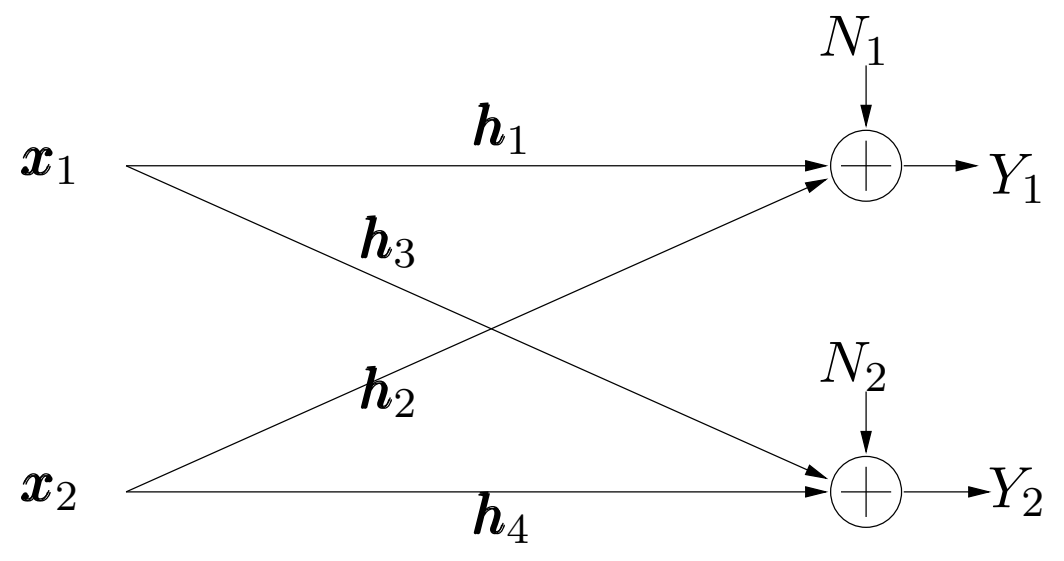

Fig. 1. The two-user Gaussian MISO IC.

Problem (2) is a non-convex optimization problem. For each $\mu_1$ and $\mu_2$ pair, all possible $\mathbf{S}_1$ and $\mathbf{S}_2$ must be exhausted over to find the solution of problem (2). To obtain the entire SUD rate region, one has to go through this exhaustive search for all the $\mu_1$ and $\mu_2$ pairs. This exhaustive search is prohibitive when $t_1$ and $t_2$ become large.

In the following, we first convert problem (2) into a family of convex optimization problems, and then obtain their closed-form solutions.

### A. Problem reformulation

We first define the following optimization problems:

$$\begin{aligned} \max \quad & \boldsymbol{h}_1^\dagger \mathbf{S}_1 \boldsymbol{h}_1 \\ \text{subject to} \quad & \boldsymbol{h}_3^\dagger \mathbf{S}_1 \boldsymbol{h}_3 = z_1^2 \\ & \text{tr}\left(\mathbf{S}_1\right) \leq P_1, \quad \mathbf{S}_1 \succeq 0, \end{aligned} \tag{4}$$

and

$$\begin{aligned} \max \quad & \boldsymbol{h}_4^\dagger \mathbf{S}_2 \boldsymbol{h}_4 \\ \text{subject to} \quad & \boldsymbol{h}_2^\dagger \mathbf{S}_2 \boldsymbol{h}_2 = z_2^2 \\ & \text{tr}\left(\mathbf{S}_2\right) \leq P_2, \quad \mathbf{S}_2 \succeq 0. \end{aligned} \tag{5}$$

In order for problems (4) and (5) to be feasible, we require

$$0 \leq z_1 \leq \max_{\text{tr}(\mathbf{S}_1) \leq P_1, \mathbf{S}_1 \succeq \mathbf{0}} \sqrt{\boldsymbol{h}_3^\dagger \mathbf{S}_1 \boldsymbol{h}_3} = \sqrt{P_1} \|\boldsymbol{h}_3\| \tag{6}$$

$$\text{and} \quad 0 \leq z_2 \leq \max_{\text{tr}(\mathbf{S}_2) \leq P_2, \mathbf{S}_2 \succeq \mathbf{0}} \sqrt{\boldsymbol{h}_2^\dagger \mathbf{S}_2 \boldsymbol{h}_2} = \sqrt{P_2} \|\boldsymbol{h}_2\|. \tag{7}$$

We now establish the equivalence between problem (2) and the above two optimization problems.

*Lemma 1:* For any non-negative scalars $\mu_1$ and $\mu_2$, the optimal solution $\mathbf{S}_1^*$ and $\mathbf{S}_2^*$ for problem (2) is also an optimal solution for problems (4) and (5) with $z_1^2 = z_1^{*2} = \boldsymbol{h}_3^\dagger \mathbf{S}_1^* \boldsymbol{h}_3$ and $z_2^2 = z_2^{*2} = \boldsymbol{h}_2^\dagger \mathbf{S}_2^* \boldsymbol{h}_2$.

*Proof:* Problem (2) is equivalent to the following optimization problem for the same $\mu_1$ and $\mu_2$:

$$\begin{aligned} \max \quad & \left[ \mu_1 \log\left(1 + \frac{\boldsymbol{h}_1^\dagger \mathbf{S}_1 \boldsymbol{h}_1}{1 + z_2^{*2}}\right) + \mu_2 \log\left(1 + \frac{\boldsymbol{h}_4^\dagger \mathbf{S}_2 \boldsymbol{h}_4}{1 + z_1^{*2}}\right) \right] \\ \text{subject to} \quad & \boldsymbol{h}_3^\dagger \mathbf{S}_1 \boldsymbol{h}_3 = z_1^{*2}, \quad \boldsymbol{h}_2^\dagger \mathbf{S}_2 \boldsymbol{h}_2 = z_2^{*2}, \\ & \text{tr}\left(\mathbf{S}_i\right) \leq P_i, \quad \mathbf{S}_i \succeq 0, \quad i = 1, 2. \end{aligned} \tag{8}$$

The equivalence is established as follows. First, the maximum of problem (2) is no smaller than that of problem (8), since problem (8) has extra constraints $\boldsymbol{h}_3^\dagger \mathbf{S}_1 \boldsymbol{h}_3 = z_1^{*2}$ and $\boldsymbol{h}_2^\dagger \mathbf{S}_2 \boldsymbol{h}_2 = z_2^{*2}$. On the other hand, the maximum of problem (2) is no greater than that of problem (8), since $\mathbf{S}_1^*$ and $\mathbf{S}_2^*$ are also feasible for problem (8), which are the optimal solutions for problem (2). Therefore, problems (2) and (8) are equivalent. We now recognize that problem (8) is equivalent to problems (4) and (5) with $z_1 = z_1^*$ and $z_2 = z_2^*$, which can be solved individually. ■

We remark that the optimization problem (8) can not be solved independently as the constraint parameters $z_1$ and $z_2$ depend on the unknown optimal covariances. That is, unless the optimal $\mathbf{S}_1^*$ and $\mathbf{S}_2^*$ of problem (2) are obtained, the equivalent optimization problems in the form of (4) and (5) cannot be parameterized. However, this problem reformulation becomes especially powerful when we need to find the entire achievable rate region (or its boundary points) and to study the optimal signaling structure. Even though one cannot solve any individual optimization problem (2) by the corresponding problem (8), Lemma 1 establishes the following crucial fact that enables us to obtain the entire SUD rate region without explicitly solving (8):

$$\bigcup_{\text{all } \mu_1, \mu_2} \{\mathbf{S}_1^*(\mu_1, \mu_2), \mathbf{S}_2^*(\mu_1, \mu_2)\} \subseteq \bigcup_{\text{all } z_1, z_2} \left\{\bar{\mathbf{S}}_1^*(z_1, z_2), \bar{\mathbf{S}}_2^*(z_1, z_2)\right\}, \tag{9}$$

where the left-hand side denotes the collection of all the optimal solutions of problem (2) found by exhausting over $\mu_1$ and $\mu_2$, and the right-hand side denotes the collection of all the optimal solutions of problems (4) and (5) found by exhausting over $z_1$ and $z_2$. Since the SUD rate region is determined by the left-hand side of (9), Lemma 1 successfully converts a family of non-convex optimization problems (2) into a family of equivalent convex optimization problems (4) and (5).

To be more precise, instead of solving the family of non-convex optimization problems by exhausting over $\mu_1$ and $\mu_2$, one can instead solve the family of convex optimization problems by exhausting $z_1$ and $z_2$ over the range specified by (6) and (7). We now proceed to obtain closed-form solutions for problems (4) and (5).

### *B. Optimal Solution*

By symmetry, we need only solve problem (4). Assume the singular-value decomposition (SVD) of $\boldsymbol{h}_i$ is

$$\boldsymbol{h}_i = \mathbf{U}_i \begin{bmatrix} \|\boldsymbol{h}_i\| \\ \mathbf{0} \end{bmatrix}, \quad i = 1, \ldots, 4, \tag{10}$$

where $\mathbf{U}_i\mathbf{U}_i^\dagger = \mathbf{I}$. Define

$$\widehat{\boldsymbol{h}}_1 = \mathbf{U}_3^\dagger \boldsymbol{h}_1 = \begin{bmatrix} \widehat{h}_{11} \\ \boldsymbol{\beta} \end{bmatrix}, \tag{11}$$

where $\widehat{h}_{11} = \left(\widehat{\boldsymbol{h}}_1\right)_1$ and $\boldsymbol{\beta}$ is a $(t_1 - 1) \times 1$ vector. We have the following lemma.

*Lemma 2:* Assuming the optimization problem (4) is feasible, the following $\mathbf{S}_1^*$'s are optimal:

- If $\boldsymbol{h}_1$ and $\boldsymbol{h}_3$ are linearly independent (consequently $\|\boldsymbol{h}_3\| \neq 0$, and $\|\boldsymbol{\beta}\| \neq 0$), then

$$\mathbf{S}_1^* = \boldsymbol{\gamma}_1 \boldsymbol{\gamma}_1^\dagger, \tag{12}$$

  and the achieved maximum is

$$\boldsymbol{h}_1^\dagger \mathbf{S}_1^* \boldsymbol{h}_1 = \left( \frac{z_1 \left| \boldsymbol{h}_3^\dagger \boldsymbol{h}_1 \right|}{\|\boldsymbol{h}_3\|^2} + \sqrt{\left( \|\boldsymbol{h}_1\|^2 - \frac{1}{\|\boldsymbol{h}_3\|^2} \left| \boldsymbol{h}_3^\dagger \boldsymbol{h}_1 \right|^2 \right) \left( P_1 - \frac{z_1^2}{\|\boldsymbol{h}_3\|^2} \right)} \right)^2, \tag{13}$$

  where

$$\boldsymbol{\gamma}_1 = \mathbf{U}_3 \begin{bmatrix} \dfrac{z_1}{\|\boldsymbol{h}_3\|} \\ k\sqrt{P_1 - \dfrac{z_1^2}{\|\boldsymbol{h}_3\|^2}} \cdot \dfrac{\boldsymbol{\beta}}{\|\boldsymbol{\beta}\|} \end{bmatrix} \tag{14}$$

$$\text{and} \quad k = \begin{cases} \dfrac{\widehat{h}_{11}^\dagger}{\left|\widehat{h}_{11}\right|}, & \text{if } \widehat{h}_{11} \neq 0 \\ 1, & \text{otherwise.} \end{cases} \tag{15}$$

- If $\boldsymbol{h}_1$ and $\boldsymbol{h}_3$ are linearly dependent and $\|\boldsymbol{h}_3\| \neq 0$, then

$$\mathbf{S}_1^* = \frac{z_1^2}{\|\boldsymbol{h}_3\|^2} \boldsymbol{h}_3 \boldsymbol{h}_3^\dagger \tag{16}$$

$$\text{and} \quad \boldsymbol{h}_1^\dagger \mathbf{S}_1^* \boldsymbol{h}_1 = \frac{z_1^2}{\|\boldsymbol{h}_3\|^2} \left| \boldsymbol{h}_1^\dagger \boldsymbol{h}_3 \right|^2. \tag{17}$$

- If $\|\boldsymbol{h}_3\| = 0$ (hence $z_1 = 0$), then

$$\mathbf{S}_1^* = \frac{P_1}{\|\boldsymbol{h}_1\|^2} \boldsymbol{h}_1 \boldsymbol{h}_1^\dagger, \tag{18}$$

$$\text{and} \quad \boldsymbol{h}_1^\dagger \mathbf{S}_1^* \boldsymbol{h}_1 = P_1 \|\boldsymbol{h}_1\|^2. \tag{19}$$

Moreover, for all above cases, we have

$$\text{rank}\left(\mathbf{S}_1^*\right) \leq 1. \tag{20}$$

The proof is given in Appendix A.

Lemma 2 shows that, for a fixed interference power $z_1^2$, transmitter beamforming maximizes the received signal power[1]. If $\boldsymbol{h}_1$ and $\boldsymbol{h}_3$ are linearly independent, the quadratic constraint of (4) defines a set of beamforming vectors whose projections on $\boldsymbol{h}_3$ have equal length $z_1$. Among all these vectors, the one that has the largest length of the projection on $\boldsymbol{h}_1$ is the optimal beamforming vector.

### *C. The SUD rate region of a two-user MISO IC*

With Lemmas 1 and 2, we obtain the SUD rate region of a two-user MISO IC.

*Theorem 1:* The SUD rate region of a two-user MISO IC with complex channels is

$$\bigcup_{\substack{\psi_1\in\left[0,\frac{\pi}{2}-\theta_1\right]\\ \psi_2\in\left[0,\frac{\pi}{2}-\theta_2\right]}} \left\{ \begin{array}{l} R_1 \leq \log\left(1+\dfrac{P_1\|\boldsymbol{h}_1\|^2\sin^2(\theta_1+\psi_1)}{1+P_2\|\boldsymbol{h}_2\|^2\sin^2\psi_2}\right) \\ R_2 \leq \log\left(1+\dfrac{P_2\|\boldsymbol{h}_4\|^2\sin^2(\theta_2+\psi_2)}{1+P_1\|\boldsymbol{h}_3\|^2\sin^2\psi_1}\right) \end{array} \right\}, \tag{21}$$

where

$$\theta_1 = \begin{cases} \cos^{-1}\dfrac{\left|\boldsymbol{h}_3^\dagger\boldsymbol{h}_1\right|}{\|\boldsymbol{h}_1\|\cdot\|\boldsymbol{h}_3\|} & \text{if } \|\boldsymbol{h}_1\|\neq 0, \quad \|\boldsymbol{h}_3\|\neq 0 \\ 0 & \text{otherwise,} \end{cases}$$

$$\text{and} \quad \theta_2 = \begin{cases} \cos^{-1}\dfrac{\left|\boldsymbol{h}_2^\dagger\boldsymbol{h}_4\right|}{\|\boldsymbol{h}_2\|\cdot\|\boldsymbol{h}_4\|} & \text{if } \|\boldsymbol{h}_2\|\neq 0, \quad \|\boldsymbol{h}_4\|\neq 0 \\ 0 & \text{otherwise.} \end{cases}$$

Furthermore, the boundary points of the rate region can be achieved by restricting each transmitter to implement beamforming.

*Proof:* We first assume that $\boldsymbol{h}_1$ is linearly independent of $\boldsymbol{h}_3$, and that $\boldsymbol{h}_2$ is linearly independent of $\boldsymbol{h}_4$. Define

$$\phi_1(z_1) = \tan^{-1}\sqrt{\frac{z_1^2}{P_1\|\boldsymbol{h}_3\|^2 - z_1^2}};$$

then (13) becomes

$$\boldsymbol{h}_1^\dagger \mathbf{S}_1^* \boldsymbol{h}_1 = P_1\|\boldsymbol{h}_1\|^2\sin^2\left(\theta_1+\phi_1(z_1)\right).$$

Similarly, the maximum of problem (5) is

$$\boldsymbol{h}_4^\dagger \mathbf{S}_2^* \boldsymbol{h}_4 = P_2\|\boldsymbol{h}_4\|^2\sin^2\left(\theta_2+\phi_2(z_2)\right),$$

[1]In Appendix A, we show that when $\boldsymbol{h}_1^\dagger\boldsymbol{h}_3 = 0$, there exist some matrices that are not beamforming matrices but still maximize problem (4). This also happens when $\boldsymbol{h}_1$ and $\boldsymbol{h}_3$ are linearly dependent. However, these does not contradict our conclusion that beamforming is optimal.

where

$$\phi_2(z_2) = \tan^{-1}\sqrt{\frac{z_2^2}{P_2\|\boldsymbol{h}_2\|^2 - z_2^2}}.$$

Therefore, the achievable rate region determined by problem (2) is

$$\bigcup_{\substack{z_1^2\in[0,P_1\|\boldsymbol{h}_3\|^2]\\ z_2^2\in[0,P_2\|\boldsymbol{h}_2\|^2]}} \left\{ \begin{array}{l} R_1 \leq \log\left(1+\dfrac{P_1\|\boldsymbol{h}_1\|^2\sin^2\left[\theta_1+\phi_1(z_1)\right]}{1+z_2^2}\right) \\ R_2 \leq \log\left(1+\dfrac{P_2\|\boldsymbol{h}_4\|^2\sin^2\left[\theta_2+\phi_2(z_2)\right]}{1+z_1^2}\right) \end{array} \right\}. \tag{22}$$

On defining $\psi_1 = \tan^{-1}\sqrt{\frac{z_1^2}{P_1\|\boldsymbol{h}_3\|^2 - z_1^2}}$ and $\psi_2 = \tan^{-1}\sqrt{\frac{z_2^2}{P_2\|\boldsymbol{h}_2\|^2 - z_2^2}}$, the interference power and the useful signal power caused by transmitter 1 are given respectively by

$$z_1^2 = P_1\|\boldsymbol{h}_3\|^2\sin^2\psi_1, \tag{23}$$

and

$$\boldsymbol{h}_1^\dagger \mathbf{S}_1^* \boldsymbol{h}_1 = P_1\|\boldsymbol{h}_1\|^2\sin^2(\theta_1+\psi_1). \tag{24}$$

Similarly, the interference power and the useful signal power caused by transmitter 2 are given respectively by

$$z_2^2 = P_2\|\boldsymbol{h}_2\|^2\sin^2\psi_2, \tag{25}$$

and

$$\boldsymbol{h}_4^\dagger \mathbf{S}_2^* \boldsymbol{h}_4 = P_2\|\boldsymbol{h}_4\|^2\sin^2(\theta_2+\psi_2). \tag{26}$$

Since $\psi_1$ varies continuously in $\left[0, \frac{\pi}{2}\right]$ as $z_1$ varies in $\left[0, P\|\boldsymbol{h}_3\|^2\right]$ (and similar for $\psi_2$), the region in (22) is the same as

$$\bigcup_{\substack{\psi_1\in\left[0,\frac{\pi}{2}\right]\\ \psi_2\in\left[0,\frac{\pi}{2}\right]}} \left\{ \begin{array}{l} R_1 \leq \log\left(1+\dfrac{P_1\|\boldsymbol{h}_1\|^2\sin^2\left(\theta_1+\psi_1\right)}{1+P_2\|\boldsymbol{h}_2\|^2\sin^2\psi_2}\right) \\ R_2 \leq \log\left(1+\dfrac{P_2\|\boldsymbol{h}_4\|^2\sin^2\left(\theta_2+\psi_2\right)}{1+P_1\|\boldsymbol{h}_3\|^2\sin^2\psi_1}\right) \end{array} \right\}. \tag{27}$$

When $\psi_i \in \left[\frac{\pi}{2}-\theta_i, \frac{\pi}{2}\right]$, the useful signal power $P_1\|\boldsymbol{h}_1\|^2\sin^2(\theta_1+\psi_1)$ or $P_2\|\boldsymbol{h}_4\|^2\sin^2(\theta_2+\psi_2)$ decreases as the interference power $P_1\|\boldsymbol{h}_3\|^2\sin^2\psi_1$ or $P_2\|\boldsymbol{h}_2\|^2\sin^2\psi_2$ increases. As such, the rate pairs associated with $\psi_i \in \left[\frac{\pi}{2}-\theta_i, \frac{\pi}{2}\right]$ are interior points of the set (27). Therefore, (27) can be simplified into (21).

In the cases where $\boldsymbol{h}_1$ and $\boldsymbol{h}_3$ are linearly dependent, (21) is still the SUD rate region. This is due to the fact that (17) can be expressed in (24) when $\theta_1 = 0$; and (19) can also be expressed in (24) when $\theta_1 = 0$ and $\psi_1 = \frac{\pi}{2}$.

For each choice of $\psi_1$ and $\psi_2$, the corresponding $\mathbf{S}_1^*$ can be obtained from (12), (16) or (18), and similarly for $\mathbf{S}_2^*$. ■

The rate region in (21) is characterized by $\psi_1$ and $\psi_2$ which also determine the interference powers (23) and (25) at the two receivers. Compared to the original problem (2) which is characterized by the slopes $(-\frac{\mu_1}{\mu_2})$ of the boundary points and requires the solution of a family of non-convex optimizations, Theorem 1 gives closed-form solutions by problem reformulation. Moreover, Theorem 1 shows that to achieve the rate pairs on the boundary of the SUD rate region, the transmitter can restrict itself to a simple beamforming strategy.

In the SUD rate region (21), we point out several special rate pairs.

- $\psi_1 = \psi_2 = 0$. This corresponds to a zero-forcing (ZF) beamforming rate pair as both transmitters generate no interference to the unintended receiver. Thus $R_1 = \log\left(1 + P_1\|\boldsymbol{h}_1\|^2 \sin^2\theta_1\right)$ and $R_2 = \log\left(1 + P_2\|\boldsymbol{h}_4\|^2 \sin^2\theta_2\right)$ are the maximum ZF beamforming rates. In general, this rate pair is in the interior of the SUD rate region.
- $\psi_1 = 0$, $\psi_2 = \frac{\pi}{2} - \theta_2$. This case shows that user 1 can communicate at a rate no greater than $R_1 = \log\left(1 + P_1\|\boldsymbol{h}_1\|^2 \sin^2\theta_1 / \left(1 + P_2\|\boldsymbol{h}_2\|^2\cos^2\theta_2\right)\right)$ when user 2 is at the maximum rate $R_2 = \log\left(1 + P_2\|\boldsymbol{h}_4\|^2\right)$. This corresponds to a corner point on the rate region.
- $\psi_1 = \frac{\pi}{2} - \theta_1$, $\psi_2 = 0$. This is the other corner point of the rate region. User 2 can communicate at a rate no greater than $R_2 = \log\left(1 + P_2\|\boldsymbol{h}_4\|^2 \sin^2\theta_2 / \left(1 + P_1\|\boldsymbol{h}_3\|^2\cos^2\theta_1\right)\right)$ when user 1 is at the maximum rate $R_1 = \log\left(1 + P_1\|\boldsymbol{h}_1\|^2\right)$.

When $\boldsymbol{h}_1$ and $\boldsymbol{h}_3$, and $\boldsymbol{h}_2$ and $\boldsymbol{h}_4$ are respectively linearly independent, both transmitters use all their power to achieve the largest rates. However, if either of the above two vector pairs is linearly dependent, the transmitters do not necessarily use all the power. An example is the scalar Gaussian IC.

*Lemma 3:* [28, Theorem 6] If $\boldsymbol{h}_1 = \boldsymbol{h}_4 = 1$, $\boldsymbol{h}_2 = \sqrt{a}$ and $\boldsymbol{h}_3 = \sqrt{b}$, then the maximum SUD sum rate is

$$R_s = \max\left\{f(P_1, P_2), f(0, P_2), f(P_1, 0)\right\}, \tag{28}$$

where

$$f(p_1, p_2) = \log\left(1 + \frac{p_1}{1 + ap_2}\right) + \log\left(1 + \frac{p_2}{1 + bp_1}\right).$$

Therefore, the maximum SUD sum rate for a scalar IC is achieved by letting both users use all the power, or letting one user use all the power while keeping the other user silent. The contrast between a scalar IC and a MISO IC with linearly independent channels is largely due to the existence of and the interplay between the spatial diversity and multi-user diversity of a MISO IC. There is a tradeoff between the power of the intended signal at its own receiver and the interference power at the other receiver. For a scalar Gaussian IC, these two signals overlap in the same subspace and are proportional to each other, i.e., the channels are always linearly dependent. An increase of the intended signal power always results in an increase of the interference power (see (23) and (24) when $\theta_1$ is $0$). The optimal tradeoff is achieved by choosing the appropriate power at the transmitters. As shown in (28), the optimal tradeoff does not necessarily require that both users use all the power. But for the MISO IC with linearly independent channels, the intended link and the interference link are in non-overlapping subspaces. Therefore, the optimal tradeoff is achieved by choosing the optimal beamforming subspaces while using all the power.

Fig. 2 is an illustration of the beamforming vector of MISO IC. For simplicity, the channel vectors $\mathbf{h}_1$ and $\mathbf{h}_3$ are assumed to be $2 \times 1$ real vectors with unit lengths. The angle between $\mathbf{h}_1$ and $\mathbf{h}_2$ is $\theta$. The disc with radius $\sqrt{P}$ contains all possible beamforming vectors that satisfy the power constraint. $E$ and $F$ are on the circle, the projections of vectors $\overline{OE}$ and $\overline{OF}$ on $\mathbf{h}_3$ both have length $z$. Then all the vectors on the line segment $EF$ satisfy the power constraint $P$ and the interference constraint $z$. Among those vectors, $\overline{OE}$ has the greatest length of projection on $\mathbf{h}_1$. Therefore, $\overline{OE}$ is the optimal beamforming vector $\boldsymbol{\gamma}$ given the interference constraint $z$. It can be shown that the angle between $\overline{OE}$ and $\overline{EF}$ is $\psi$ and the length of the projection of $\overline{OE}$ on $\mathbf{h}_1$ is $\sqrt{P}\sin(\theta + \psi)$.

The reduction of the non-convex optimization problem (2) to the equivalent optimization problem (4) is obtained by fixing the interference power while maximizing the useful signal power. This method is equivalent to fixing the useful signal power while minimizing the interference power. This requires solving the following optimization problems:

$$\begin{aligned} \min \quad & \boldsymbol{h}_3^\dagger \mathbf{S}_1 \boldsymbol{h}_3 \\ \text{subject to} \quad & \boldsymbol{h}_1^\dagger \mathbf{S}_1 \boldsymbol{h}_1 = z_3^2, \\ & \text{tr}\,(\mathbf{S}_1) \le P_1, \quad \mathbf{S}_1 \succeq \mathbf{0}, \end{aligned}$$

and

$$\begin{aligned} \min \quad & \boldsymbol{h}_2^\dagger \mathbf{S}_2 \boldsymbol{h}_2 \\ \text{subject to} \quad & \boldsymbol{h}_4^\dagger \mathbf{S}_2 \boldsymbol{h}_4 = z_4^2. \\ & \text{tr}\,(\mathbf{S}_2) \le P_2, \quad \mathbf{S}_2 \succeq \mathbf{0}. \end{aligned}$$

The above two problems can be solved in the same way as problem (4), and the rate region can be similarly obtained. For these two proposed methods, the constraints are imposed either on the interference powers or the useful signal powers. One can also combine these two methods. For example, we can impose a constraint on the interference power caused by transmitter $1$ while maximizing the useful power for receiver $1$, and in the meantime, impose a constraint on the signal power on receiver $2$ while minimizing the interference power caused by transmitter $2$.

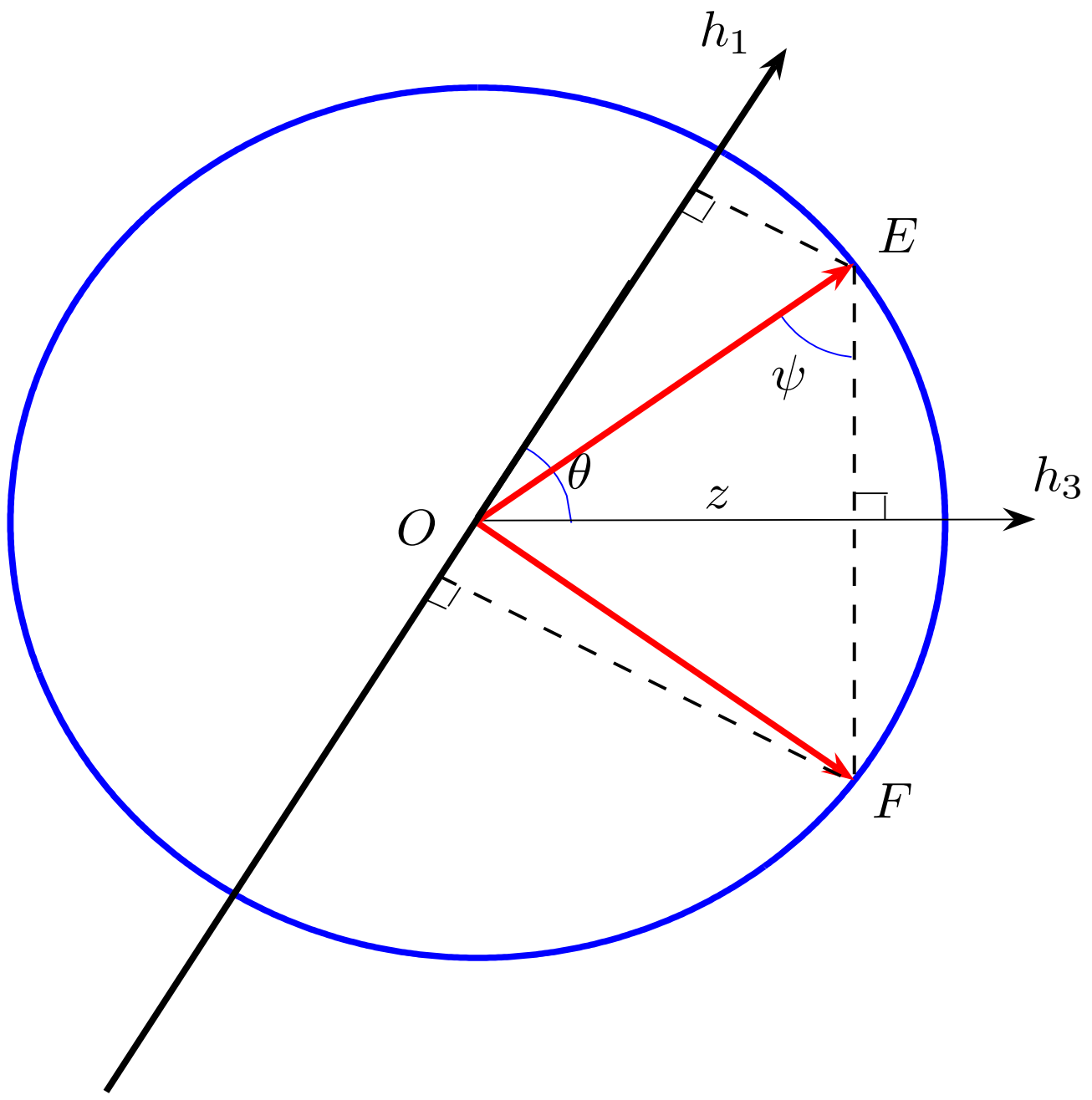

Fig. 2. A geometric explanation of beamforming in the MISO IC.

### D. Interference-limited SUD rate region

As Theorem 1 is obtained by examining the relationship between the interference power and the useful signal power, we can easily apply it to the MISO IC under interference power constraints.

*Theorem 2:* For the MISO IC defined in (1) with $\|\boldsymbol{h}_2\| \neq 0$, $\|\boldsymbol{h}_3\| \neq 0$, and with two additional constraints $\boldsymbol{h}_3^\dagger \mathbf{S}_1 \boldsymbol{h}_3 \leq Q_1$ and $\boldsymbol{h}_2^\dagger \mathbf{S}_2 \boldsymbol{h}_2 \leq Q_2$ on the interference powers, the SUD rate region is

$$\bigcup_{\substack{\psi_1 \in [0, \bar{\theta}_1] \\ \psi_2 \in [0, \bar{\theta}_2]}} \left\{ \begin{array}{l} R_1 \leq \log\left(1 + \dfrac{P_1 \|\boldsymbol{h}_1\|^2 \sin^2(\theta_1 + \psi_1)}{1 + P_2 \|\boldsymbol{h}_2\|^2 \sin^2 \psi_2}\right) \\ R_2 \leq \log\left(1 + \dfrac{P_2 \|\boldsymbol{h}_4\|^2 \sin^2(\theta_2 + \psi_2)}{1 + P_1 \|\boldsymbol{h}_3\|^2 \sin^2 \psi_1}\right) \end{array} \right\}, \tag{29}$$

where

$$\bar{\theta}_1 = \min\left\{\frac{\pi}{2} - \theta_1, \quad \sin^{-1}\sqrt{\frac{Q_1}{P_1\|\boldsymbol{h}_3\|^2}}\right\}$$

$$\text{and} \quad \bar{\theta}_2 = \min\left\{\frac{\pi}{2} - \theta_2, \quad \sin^{-1}\sqrt{\frac{Q_2}{P_2\|\boldsymbol{h}_2\|^2}}\right\}.$$

The proof is straightforward from Theorem 1.

When $Q_1 \geq P_1\|\boldsymbol{h}_3\|^2\cos^2\theta_1$ and $Q_2 \geq P_2\|\boldsymbol{h}_2\|^2\cos^2\theta_2$, (29) is exactly (21). Therefore, when the interference constraints are larger than the above thresholds, these constraints do not change the SUD rate region. Another extreme case is that in which neither user is allowed to generate interference to the other user. This is also the ZF rate region which is a rectangle determined by $Q_1 = Q_2 = 0$.

## III. Multi-user MISO IC with single-user detection

In this section, we generalize our study of the two-user case to the general multi-user MISO IC. The key is, again, the problem reformulation as illustrated in Lemma 1 for the two-user case. For the general $m$-user MISO IC, we prove the optimality of beamforming with an SUD receiver. We then given the explicit description of the SUD rate region for a three-user MISO IC, and generalize it to the $m$-user case.

### *A. Optimality of beamforming for an $m$-user MISO IC*

Define the received signal of the $i$th user, $i = 1, \cdots, m$, as

$$Y_i = \sum_{j=1}^{m} \boldsymbol{h}_{ji}^{\dagger}\boldsymbol{x}_j + N_i, \tag{30}$$

where $\boldsymbol{x}_j$ is the $t_j \times 1$ transmitted signal vector of user $j$, $\boldsymbol{h}_{ji}$ is the $t_j \times 1$ complex channel vector from the $j$th transmitter to the $i$th receiver, and $N_i$ is unit variance circularly symmetric complex Gaussian noise. The power constraint for user $i$ is $\operatorname{tr}(\mathbf{S}_i) \leq P_i$ where $\mathbf{S}_i = E\left[\boldsymbol{x}_i\boldsymbol{x}_i^{\dagger}\right]$. As with the two-user case, the input signals $\boldsymbol{x}_j$'s are all zero-mean Gaussian vectors and each receiver treats interference as noise.

Lemma 1 can be easily extended to multi-user MISO ICs, as follows.

*Lemma 4:* For any vector $\boldsymbol{\mu} = [\mu_1, \cdots, \mu_m]$ with non-negative components, the optimal solution $[\mathbf{S}_1^*, \cdots, \mathbf{S}_m^*]$ for the following optimization problem:

$$\begin{aligned} \max \quad & \sum_{i=1}^{m} \mu_i R_i \\ \text{subject to} \quad & R_i = \log\left(1 + \frac{\boldsymbol{h}_{ii}^{\dagger}\mathbf{S}_i\boldsymbol{h}_{ii}}{1 + \sum_{j=1, j\neq i}^{n}\boldsymbol{h}_{ji}^{\dagger}\mathbf{S}_j\boldsymbol{h}_{ji}}\right) \\ & \operatorname{tr}(\mathbf{S}_i) \leq P_i, \quad \mathbf{S}_i \succeq \mathbf{0}, \quad i = 1, \cdots m, \end{aligned} \tag{31}$$

is also an optimal solution for the problem

$$\begin{aligned} \max \quad & \boldsymbol{h}_{ii}^{\dagger}\mathbf{S}_i\boldsymbol{h}_{ii} \\ \text{subject to} \quad & \boldsymbol{h}_{ij}^{\dagger}\mathbf{S}_i\boldsymbol{h}_{ij} = z_{ij}^2, \\ & \text{tr}(\mathbf{S}_i) \le P_i, \quad \mathbf{S}_i \succeq 0 \\ & i,j = 1,\ldots,m, \quad i \neq j, \end{aligned} \tag{32}$$

with $z_{ij}^2 = z_{ij}^{*2} = \boldsymbol{h}_{ij}^{\dagger}\mathbf{S}_i^*\boldsymbol{h}_{ij}$.

Following the same problem reformulation procedure used in Section II, to characterize the SUD rate region of an $m$-user MISO IC, the key appears to be the solution of (32) where $z_{ij}^2$ is a preselected constant denoting the interference power at the $j$th receiver caused by the $i$th transmitter, and $\mathbf{S}_i$ and $P_i$ are the covariance matrix and power constraint for the $i$th transmitter.

Unlike problem (4), the optimal covariance matrix for (32) with any given $z_{ij}$'s need not necessarily be a beamforming matrix. Here is an example.

*Example 1:* Consider the channels

$$\begin{aligned} & \boldsymbol{h}_{11} = [1.9574, 0.5045, 1.8645, -0.3398]^T, \\ & \boldsymbol{h}_{12} = [-1.1398, -0.2111, 1.1902, -1.1162]^T \\ \text{and} \quad & \boldsymbol{h}_{13} = [0.6353, -0.6014, 0.5512, -1.0998]^T. \end{aligned}$$

The optimal covariance matrix with the constraints $P = 1$, $z_{12}^2 = 0.3$ and $z_{13}^2 = 0.6$ is

$$\mathbf{S}_1^* = \begin{bmatrix} 0.3784 & 0.1355 & 0.4103 & -0.0461 \\ 0.1355 & 0.0499 & 0.1587 & -0.0220 \\ 0.4103 & 0.1587 & 0.5443 & -0.0965 \\ -0.0461 & -0.0220 & -0.0965 & 0.0273 \end{bmatrix}, \quad \text{rank}(\mathbf{S}_1^*) = 2.$$

By restricting to beamforming, the optimal covariance matrix is

$$\bar{\mathbf{S}}_1 = \begin{bmatrix} 0.2320 & 0.1198 & 0.3985 & -0.0711 \\ 0.1198 & 0.0619 & 0.2058 & -0.0367 \\ 0.3985 & 0.2058 & 0.6843 & -0.1220 \\ -0.0711 & -0.0367 & -0.1220 & 0.0218 \end{bmatrix}, \quad \text{rank}(\bar{\mathbf{S}}_1) = 1.$$

We have

$$\boldsymbol{h}_1^{\dagger}\mathbf{S}_1^*\boldsymbol{h}_1 = 7.1100 > \boldsymbol{h}_1^{\dagger}\bar{\mathbf{S}}_1\boldsymbol{h}_1 = 7.0805.$$

However, the above example does not mean that beamforming is not optimal for the SUD rate region of an $m$-user MISO IC. The reason is that the optimization problem (32) requires the interference powers to be exactly $z_{ij}^2$. Without knowing the values of all ${z_{ij}^*}^2$, exhausting over all $z_{ij}^2$ will result in some rate pairs not on the boundaries of the SUD rate region. As we intend to establish the optimality of beamforming for achieving the boundary points of the rate region, we resort instead to the following more general formulation, i.e., the interference power is bounded by $z_{ij}^2$, namely

$$\begin{aligned} \max \quad & \boldsymbol{h}_{ii}^\dagger \mathbf{S}_i \boldsymbol{h}_{ii} \\ \text{subject to} \quad & \boldsymbol{h}_{ij}^\dagger \mathbf{S}_i \boldsymbol{h}_{ij} \leq z_{ij}^2, \\ & \text{tr}(\mathbf{S}_i) \leq P_i, \quad \mathbf{S}_i \succeq 0 \\ & i, j = 1, \ldots, m, \quad i \neq j, \end{aligned} \tag{33}$$

This modified problem has its maximum that is no smaller than that of problem (32) for any values of $z_{ij}^2$ because of the relaxed constraint on the interference power. Therefore, if we can prove that beamforming is optimal for (33), then beamforming must be optimal for problem (31) even if it may not be optimal for (32). Such a strategy has been used in Theorem 1 where we let $\psi_i$ vary in $\left[0, \frac{\pi}{2} - \theta_i\right]$ instead of the entire interval $\left[0, \frac{\pi}{2}\right]$ because only in the specified interval does the useful signal power increase as the interference power increases. Based on the modified optimization problem (33), we obtain the following theorem.

*Theorem 3:* For an $m$-user MISO IC, the boundary points of the SUD rate region can be achieved by restricting each transmitter to implement beamforming.

The proof of Theorem 3 is achieved in the following steps. We first introduce Lemma 5 which allows us to solve problem (40). In the process, we establish that the rank of the entire covariance matrix can be set to the same as that of its submatrices. The final step is to show that the optimal submatrix need to have a rank no greater than 1, established via an extension of Sylvester's Law of Inertia.

We first introduce Lemma 5.

*Lemma 5:* Let $\boldsymbol{x}$ and $\boldsymbol{y}$ be two complex vectors with dimensions $t_1$ and $t_2$ respectively, and $\mathbf{K}$ be a $(t_1 + t_2) \times (t_1 + t_2)$ positive semi-definite Hermitian matrix with tr$(\mathbf{K}) \leq P$. If

$$\mathbf{K} = \begin{bmatrix} \mathbf{K}_{11} & \mathbf{K}_{21}^\dagger \\ \mathbf{K}_{21} & \mathbf{K}_{22} \end{bmatrix} \tag{34}$$

and $\mathbf{K}_{11}$ is a preselected $t_1 \times t_1$ positive semi-definite Hermitian matrix, then

$$\begin{bmatrix} \boldsymbol{x} \\ \boldsymbol{y} \end{bmatrix}^{\dagger} \mathbf{K} \begin{bmatrix} \boldsymbol{x} \\ \boldsymbol{y} \end{bmatrix} \leq \left( \sqrt{\boldsymbol{x}^{\dagger}\mathbf{K}_{11}\boldsymbol{x}} + \|\boldsymbol{y}\| \sqrt{P - \text{tr}(\mathbf{K}_{11})} \right)^2, \tag{35}$$

and the equality can be achieved by choosing $\mathbf{K} = \mathbf{K}^*$, defined as follows.

1) When $\boldsymbol{x}^{\dagger}\mathbf{K}_{11}\boldsymbol{x} \neq 0$ and $\|\boldsymbol{y}\| \neq 0$, we have

$$\mathbf{K}^* = \begin{bmatrix} \mathbf{K}_{11} & \frac{\sqrt{P - \text{tr}(\mathbf{K}_{11})}}{\|\boldsymbol{y}\| \sqrt{\boldsymbol{x}^{\dagger}\mathbf{K}_{11}\boldsymbol{x}}} \mathbf{K}_{11}\boldsymbol{x}\boldsymbol{y}^{\dagger} \\ \frac{\sqrt{P - \text{tr}(\mathbf{K}_{11})}}{\|\boldsymbol{y}\| \sqrt{\boldsymbol{x}^{\dagger}\mathbf{K}_{11}\boldsymbol{x}}} \boldsymbol{y}\boldsymbol{x}^{\dagger}\mathbf{K}_{11} & \frac{P - \text{tr}(\mathbf{K}_{11})}{\|\boldsymbol{y}\|^2} \boldsymbol{y}\boldsymbol{y}^{\dagger} \end{bmatrix}. \tag{36}$$

2) When $\boldsymbol{x}^{\dagger}\mathbf{K}_{11}\boldsymbol{x} = 0$ and $\|\boldsymbol{y}\| \neq 0$, we can set

$$\mathbf{K}^* = \begin{bmatrix} \mathbf{K}_{11} & \frac{\sqrt{P - \text{tr}(\mathbf{K}_{11})}}{\|\boldsymbol{y}\|} \mathbf{K}_{11}^{\frac{\dagger}{2}} \mathbf{1}_0 \boldsymbol{y}^{\dagger} \\ \frac{\sqrt{P - \text{tr}(\mathbf{K}_{11})}}{\|\boldsymbol{y}\|} \boldsymbol{y} \mathbf{1}_0^T \mathbf{K}_{11}^{\frac{1}{2}} & \frac{P - \text{tr}(\mathbf{K}_{11})}{\|\boldsymbol{y}\|^2} \boldsymbol{y}\boldsymbol{y}^{\dagger} \end{bmatrix}, \tag{37}$$

where

$$\mathbf{1}_0 = \begin{bmatrix} 1 \\ \mathbf{0}_{(t_1-1)\times 1} \end{bmatrix}$$

$$\text{and} \quad \mathbf{K}_{11}^{\frac{1}{2}} = \begin{bmatrix} \mathbf{\Lambda}^{\frac{1}{2}} & \mathbf{0} \\ \mathbf{0} & \mathbf{0} \end{bmatrix} \mathbf{Q},$$

with

$$\mathbf{K}_{11} = \mathbf{Q}^{\dagger} \begin{bmatrix} \mathbf{\Lambda} & \mathbf{0} \\ \mathbf{0} & \mathbf{0} \end{bmatrix} \mathbf{Q}$$

being the eigenvalue decomposition of $\mathbf{K}_{11}$, and $\mathbf{\Lambda}$ being a strictly positive diagonal matrix.

3) When $\|\boldsymbol{y}\| = 0$, we can set

$$\mathbf{K}^* = \begin{bmatrix} \mathbf{K}_{11} & \mathbf{0} \\ \mathbf{0} & \mathbf{0} \end{bmatrix}. \tag{38}$$

Moreover, for all three cases, we have

$$\text{rank}(\mathbf{K}^*) \leq \max\{\text{rank}(\mathbf{K}_{11}), 1\}. \tag{39}$$

The proof is given in Appendix B. Here are some examples of Lemma 5:

*Example 2:* Let $\boldsymbol{x}$ and $\boldsymbol{y}$ in Lemma 5 be two scalars, $\mathbf{K} \succeq \mathbf{0}$ be a $2 \times 2$ matrix with $\text{tr}(\mathbf{K}) \leq 2$.

- If $x=y=1$ and $K_{11}=1$, then from (36), we have a unique $\mathbf{K}^*=\begin{bmatrix}1 & 1\\ 1 & 1\end{bmatrix}$, and $\text{rank}(\mathbf{K}^*)=1$.
- If $x=1, y=1$ and $K_{11}=0$, then from (37) we have a unique $\mathbf{K}^*=\begin{bmatrix}0 & 0\\ 0 & 2\end{bmatrix}$, and $\text{rank}(\mathbf{K}^*)=1$.
- If $x=0, y=1$ and $K_{11}=1$, then from (37) we have $\mathbf{K}^*=\begin{bmatrix}1 & 1\\ 1 & 1\end{bmatrix}$, and $\text{rank}(\mathbf{K}^*)=1$. However, we can also choose $\mathbf{K}'=\mathbf{I}_{2\times 2}$ and we have $\text{rank}(\mathbf{K}')=2$. Therefore, the optimal $\mathbf{K}$ in this case is not unique.
- If $x=1, y=0$, and $K_{11}=1$, then from (38), we have $\mathbf{K}^*=\begin{bmatrix}1 & 0\\ 0 & 0\end{bmatrix}$, and $\text{rank}(\mathbf{K}^*)=1$. However, we can also choose $\mathbf{K}'=\mathbf{I}_{2\times 2}$ and $\text{rank}(\mathbf{K}')=2$.

It is shown from the examples that only when $\boldsymbol{x}^\dagger\mathbf{K}_{11}\boldsymbol{x}\neq 0$ and $\|\boldsymbol{y}\|\neq 0$, or $\mathbf{K}_{11}=\mathbf{0}$ and $\|\boldsymbol{y}\|\neq 0$, is the optimal $\mathbf{K}^*$ unique. Otherwise we can choose other optimal $\mathbf{K}^*$s which may not satisfy (39).

Lemma 5 is useful for the following optimization problem:

$$\begin{aligned}\max \quad & \begin{bmatrix}\boldsymbol{x}\\ \boldsymbol{y}\end{bmatrix}^\dagger \mathbf{K} \begin{bmatrix}\boldsymbol{x}\\ \boldsymbol{y}\end{bmatrix}\\ \text{subject to} \quad & h_i\left(\mathbf{K}_{11}\right)=0, \quad i=1,\cdots,n,\\ & g_j\left(\mathbf{K}_{11}\right)\leq 0, \quad j=1,\cdots,m,\\ & \text{tr}(\mathbf{K})\leq P, \quad \mathbf{K}\succeq \mathbf{0},\end{aligned} \tag{40}$$

where $h_i(\cdot)$ and $g_j(\cdot)$ are fixed functions. By Lemma 5, we can convert the above problem into

$$\begin{aligned}\max \quad & \left(\sqrt{\boldsymbol{x}^\dagger\mathbf{K}_{11}\boldsymbol{x}}+\|\boldsymbol{y}\|\sqrt{P-\text{tr}(\mathbf{K}_{11})}\right)^2\\ \text{subject to} \quad & h_i\left(\mathbf{K}_{11}\right)=0, \quad i=1,\cdots,n,\\ & g_j\left(\mathbf{K}_{11}\right)\leq 0, \quad j=1,\cdots,m,\\ & \text{tr}(\mathbf{K}_{11})\leq P, \quad \mathbf{K}\succeq \mathbf{0}.\end{aligned} \tag{41}$$

Problems (40) and (41) have the same solution. Once the optimal $\mathbf{K}_{11}$ for problem (41) is obtained, one can construct the optimal $\mathbf{K}$ for problem (40) from (36), (37) and (38). As shown by Example 2, the choices of (37) and (38) may not be unique. One can choose $\mathbf{K}$'s that are different from (37) and (38) and still achieve the same maximum.

With Lemma 5, we prove Theorem 3 as follows.

*Proof:* By symmetry, it suffices to show that for the $m$th user, the optimal covariance matrix $\mathbf{S}_m^*$ for the following optimization problem satisfies rank$(\mathbf{S}_m^*) \leq 1$:

$$\begin{aligned} \max \quad & \boldsymbol{h}_{mm}^\dagger \mathbf{S}_m \boldsymbol{h}_{mm} \\ \text{subject to} \quad & \boldsymbol{h}_{mj}^\dagger \mathbf{S}_m \boldsymbol{h}_{mj} \leq z_{mj}^2, \quad j = 1, \cdots, m-1, \\ & \text{tr}(\mathbf{S}_m) \leq P_m, \quad \mathbf{S}_m \succeq \mathbf{0}, \end{aligned} \tag{42}$$

where all the $\boldsymbol{h}_{mj}$'s are $t_m \times 1$ vectors.

We first show that problem (42) can be written as

$$\begin{aligned} \max \quad & \left( \sqrt{\boldsymbol{h}^\dagger \tilde{\mathbf{S}}_{11} \boldsymbol{h}} + \sqrt{\|\boldsymbol{h}_{mm}\|^2 - \|\boldsymbol{h}\|^2} \cdot \sqrt{P - \text{tr}(\tilde{\mathbf{S}}_{11})} \right)^2 \\ \text{subject to} \quad & \boldsymbol{h}_j^\dagger \tilde{\mathbf{S}}_{11} \boldsymbol{h}_j \leq z_{mj}^2, \quad j = 1, \cdots, m-1 \\ & \text{tr}\left(\tilde{\mathbf{S}}_{11}\right) \leq P_m, \quad \tilde{\mathbf{S}}_{11} \succeq \mathbf{0}, \end{aligned} \tag{43}$$

where $\boldsymbol{h}$ and all the $\boldsymbol{h}_j$'s, $j = 1, \cdots, m-1$, are $\bar{m} \times 1$ vectors, $\tilde{\mathbf{S}}_{11}$ is an $\bar{m} \times \bar{m}$ matrix, and $\bar{m}$ is defined as

$$\bar{m} = \min \{t_m, m-1\}. \tag{44}$$

Obviously, when $\bar{m} = t_m \leq m-1$, problem (42) is exactly the same as problem (43) if we choose $\boldsymbol{h} = \boldsymbol{h}_{mm}$, $\tilde{\mathbf{S}}_{11} = \mathbf{S}_m$ and $\boldsymbol{h}_j = \boldsymbol{h}_{mj}$. Thus, we need only show the equivalence of problems (42) and (43) when $\bar{m} = m-1 < t_m$.

Let the SVD of $\boldsymbol{h}_{m1}$ be

$$\boldsymbol{h}_{m1} = \mathbf{U}_1 \begin{bmatrix} \|\boldsymbol{h}_{m1}\| \\ \mathbf{0}_{(t_m-1)\times 1} \end{bmatrix},$$

and define

$$\mathbf{S}_m^{(1)} = \mathbf{U}_1^\dagger \mathbf{S}_m \mathbf{U}_1 \tag{45}$$

$$\text{and} \quad \mathbf{h}_{mj}^{(1)} = \mathbf{U}_1^\dagger \mathbf{h}_{mj}, \quad j = 1, \cdots, m. \tag{46}$$

Substituting (45) and (46) into (42), we obtain

$$\begin{aligned} \max \quad & \boldsymbol{h}_{mm}^{(1)\dagger}\mathbf{S}_m^{(1)}\boldsymbol{h}_{mm}^{(1)} \\ \text{subject to} \quad & \boldsymbol{h}_{mj}^{(1)\dagger}\mathbf{S}_m^{(1)}\boldsymbol{h}_{mj}^{(1)} \le z_{mj}^2, \quad j = 2, \cdots, m-1, \\ & \begin{bmatrix} \|\boldsymbol{h}_{m1}\| \\ \mathbf{0}_{(t_m-1)\times 1} \end{bmatrix}^\dagger \mathbf{S}_m^{(1)} \begin{bmatrix} \|\boldsymbol{h}_{m1}\| \\ \mathbf{0}_{(t_m-1)\times 1} \end{bmatrix} \le z_{m1}^2, \\ & \text{tr}\left(\mathbf{S}_m^{(1)}\right) \le P_m, \quad \mathbf{S}_m^{(1)} \succeq \mathbf{0}. \end{aligned} \tag{47}$$

Consider $\mathbf{h}_{m2}^{(1)}$ and let

$$\mathbf{h}_{m2}^{(1)} = \begin{bmatrix} \left(\mathbf{h}_{m2}^{(1)}\right)_1 \\ \left(\mathbf{h}_{m2}^{(1)}\right)_{2,\cdots,t_m} \end{bmatrix} = \begin{bmatrix} 1 & \mathbf{0} \\ \mathbf{0} & \mathbf{U}_2 \end{bmatrix} \begin{bmatrix} \left(\mathbf{h}_{m2}^{(1)}\right)_1 \\ \left\|\left(\mathbf{h}_{m2}^{(1)}\right)_{2,\cdots,t_m}\right\| \\ \mathbf{0}_{(t_m-2)\times 1} \end{bmatrix}, \tag{48}$$

where $\left(\mathbf{h}_{m2}^{(1)}\right)_{2,\cdots,t_m}$ is a $(t_m - 1) \times 1$ vector consisting of the second to the last elements of $\mathbf{h}_{m2}^{(1)}$. The SVD of $\left(\mathbf{h}_{m2}^{(1)}\right)_{2,\cdots,t_m}$ is

$$\left(\mathbf{h}_{m2}^{(1)}\right)_{2,\cdots,t_m} = \mathbf{U}_2 \begin{bmatrix} \left\|\left(\mathbf{h}_{m2}^{(1)}\right)_{2,\cdots,t_m}\right\| \\ \mathbf{0}_{(t_m-2)\times 1} \end{bmatrix},$$

where $\mathbf{U}_2^\dagger \mathbf{U}_2 = \mathbf{I}_{(t_m-1)\times(t_m-1)}$. Therefore

$$\begin{bmatrix} 1 & \mathbf{0} \\ \mathbf{0} & \mathbf{U}_2 \end{bmatrix}^\dagger \begin{bmatrix} 1 & \mathbf{0} \\ \mathbf{0} & \mathbf{U}_2 \end{bmatrix} = \mathbf{I}_{t_m \times t_m}.$$

Define

$$\mathbf{S}_m^{(2)} = \begin{bmatrix} 1 & \mathbf{0} \\ \mathbf{0} & \mathbf{U}_2 \end{bmatrix}^\dagger \mathbf{S}_m^{(1)} \begin{bmatrix} 1 & \mathbf{0} \\ \mathbf{0} & \mathbf{U}_2 \end{bmatrix} \tag{49}$$

$$\text{and} \quad \mathbf{h}_{mj}^{(2)} = \begin{bmatrix} 1 & \mathbf{0} \\ \mathbf{0} & \mathbf{U}_2 \end{bmatrix}^\dagger \mathbf{h}_{mj}^{(1)}, \quad j = 1, \cdots, m. \tag{50}$$

On substituting (48), (49) and (50) into (47), we have

$$\begin{aligned} \max \quad & \boldsymbol{h}_{mm}^{(2)\dagger}\mathbf{S}_m^{(2)}\boldsymbol{h}_{mm}^{(2)} \\ \text{subject to} \quad & \boldsymbol{h}_{mj}^{(2)\dagger}\mathbf{S}_m^{(2)}\boldsymbol{h}_{mj}^{(2)} \le z_{mj}^2, \quad j = 3, \cdots, m-1, \\ & \begin{bmatrix} \|\boldsymbol{h}_{m1}\| \\ \mathbf{0}_{(t_m-1)\times 1} \end{bmatrix}^\dagger \mathbf{S}_m^{(2)} \begin{bmatrix} \|\boldsymbol{h}_{m1}\| \\ \mathbf{0}_{(t_m-1)\times 1} \end{bmatrix} \le z_{m1}^2, \\ & \begin{bmatrix} \left(\mathbf{h}_{m2}^{(1)}\right)_1 \\ \left\|\left(\mathbf{h}_{m2}^{(1)}\right)_{2,\cdots,t_m}\right\| \\ \mathbf{0}_{(t_m-2)\times 1} \end{bmatrix}^\dagger \mathbf{S}_m^{(2)} \begin{bmatrix} \left(\mathbf{h}_{m2}^{(1)}\right)_1 \\ \left\|\left(\mathbf{h}_{m2}^{(1)}\right)_{2,\cdots,t_m}\right\| \\ \mathbf{0}_{(t_m-2)\times 1} \end{bmatrix} \le z_{m2}^2, \\ & \text{tr}\left(\mathbf{S}_m^{(2)}\right) \le P_m, \quad \mathbf{S}_m^{(2)} \succeq \mathbf{0}. \end{aligned} \tag{51}$$

We note that the above transformation does not change the form of the previously modified constraints (see the third lines of problems (47) and (51)). Now we continue the above procedure up to $\mathbf{h}_{m,m-1}$. In the $j$th transformation, we keep the first $j-1$ elements of $\mathbf{h}_{mj}^{(j-1)}$ and apply the SVD of the remaining $(t_m - j + 1)$ elements, and update the optimization problem. We formulate the $j$th, $j = 2, \cdots, m-1$, iteration as follows:

$$\boldsymbol{h}_{mj}^{(j-1)} = \begin{bmatrix} \left(\boldsymbol{h}_{mj}^{(j-1)}\right)_{1,\cdots,j-1} \\ \left(\boldsymbol{h}_{mj}^{(j-1)}\right)_{j,\cdots,t_m} \end{bmatrix} = \begin{bmatrix} \mathbf{I}_{(j-1)\times(j-1)} & \mathbf{0} \\ \mathbf{0} & \mathbf{U}_j \end{bmatrix} \begin{bmatrix} \left(\boldsymbol{h}_{mj}^{(j-1)}\right)_{1,\cdots,j-1} \\ \left\|\left(\boldsymbol{h}_{mj}^{(j-1)}\right)_{j,\cdots,t_m}\right\| \\ \mathbf{0}_{(t_m-j)\times 1} \end{bmatrix},$$

$$\boldsymbol{h}_{mk}^{(j)} = \begin{bmatrix} \mathbf{I}_{(j-1)\times(j-1)} & \mathbf{0} \\ \mathbf{0} & \mathbf{U}_j \end{bmatrix}^\dagger \boldsymbol{h}_{mk}^{(j-1)}, \quad k = 1, \cdots, m,$$

$$\text{and} \quad \mathbf{S}_m^{(j)} = \begin{bmatrix} \mathbf{I}_{(j-1)\times(j-1)} & \mathbf{0} \\ \mathbf{0} & \mathbf{U}_j \end{bmatrix}^\dagger \mathbf{S}_m^{(j-1)} \begin{bmatrix} \mathbf{I}_{(j-1)\times(j-1)} & \mathbf{0} \\ \mathbf{0} & \mathbf{U}_j \end{bmatrix},$$

where $\left(\boldsymbol{h}_{mj}^{(j-1)}\right)_{j,\cdots,t_m}$ denotes the $j$th to the $t_m$th elements of $\boldsymbol{h}_{mj}^{(j-1)}$, and its SVD is

$$\left(\boldsymbol{h}_{mj}^{(j-1)}\right)_{j,\cdots,t_m} = \mathbf{U}_j \begin{bmatrix} \left\|\left(\boldsymbol{h}_{mj}^{(j-1)}\right)_{j,\cdots,t_m}\right\| \\ \mathbf{0} \end{bmatrix},$$

where $\mathbf{U}_j^\dagger \mathbf{U}_j = \mathbf{I}_{(t_m-j+1)\times(t_m-j+1)}$.

Finally, we convert problem (42) into the following form:

$$\begin{aligned} \max \quad & \begin{bmatrix} \boldsymbol{h} \\ \widehat{\boldsymbol{h}} \end{bmatrix}^{\dagger} \tilde{\mathbf{S}} \begin{bmatrix} \boldsymbol{h} \\ \widehat{\boldsymbol{h}} \end{bmatrix} \\ \text{subject to} \quad & \begin{bmatrix} \boldsymbol{h}_j \\ \mathbf{0} \end{bmatrix}^{\dagger} \tilde{\mathbf{S}} \begin{bmatrix} \boldsymbol{h}_j \\ \mathbf{0} \end{bmatrix} \leq z_{mj}^2, \quad j = 1, \cdots m-1, \\ & \text{tr}\left(\tilde{\mathbf{S}}\right) \leq P_m, \quad \tilde{\mathbf{S}} \succeq \mathbf{0}, \end{aligned} \tag{52}$$

where $\boldsymbol{h}$ and $\boldsymbol{h}_j$ are $(m-1) \times 1$ vectors and $\widehat{\boldsymbol{h}}$ is a $(t_m - m + 1) \times 1$ vector. Furthermore, $\|\boldsymbol{h}_{mm}\|^2 = \|\boldsymbol{h}\|^2 + \left\|\widehat{\boldsymbol{h}}\right\|^2$. Let

$$\tilde{\mathbf{S}} = \begin{bmatrix} \tilde{\mathbf{S}}_{11} & \tilde{\mathbf{S}}_{21}^{\dagger} \\ \tilde{\mathbf{S}}_{21} & \tilde{\mathbf{S}}_{22} \end{bmatrix}, \tag{53}$$

where $\tilde{\mathbf{S}}_{11}$ is an $(m-1) \times (m-1)$ matrix. The quadratic constraints in problem (52) are

$$\begin{bmatrix} \boldsymbol{h}_j \\ \mathbf{0} \end{bmatrix}^{\dagger} \tilde{\mathbf{S}} \begin{bmatrix} \boldsymbol{h}_j \\ \mathbf{0} \end{bmatrix} = \boldsymbol{h}_j^{\dagger} \tilde{\mathbf{S}}_{11} \boldsymbol{h}_j \leq z_{mj}^2, \quad j = 1, \cdots m-1.$$

Therefore, the quadratic constraints in problem (52) are related only to $\tilde{\mathbf{S}}_{11}$. By Lemma 5, problem (52) is equivalent to problem (43).

Thus, in summary, problem (42) is equivalent to problem (43) with all the vectors in (43) being $\bar{m} \times 1$ and $\tilde{\mathbf{S}}_{11}$ being $\bar{m} \times \bar{m}$.

By Lemma 5, we can construct $\tilde{\mathbf{S}}$ in such a way that $\text{rank}\left(\tilde{\mathbf{S}}\right) \leq \max\left\{\text{rank}\left(\tilde{\mathbf{S}}_{11}\right), 1\right\}$. Let $\tilde{\mathbf{S}}_{11}^*$ be optimal for problem (43). To prove Theorem 3, it is equivalent to prove

$$\text{rank}\left(\tilde{\mathbf{S}}_{11}^*\right) \leq 1. \tag{54}$$

Furthermore, since $\tilde{\mathbf{S}}_{11}^*$ is optimal for problem (43), $\tilde{\mathbf{S}}_{11}^*$ also maximizes $\boldsymbol{h}^{\dagger}\tilde{\mathbf{S}}_{11}\boldsymbol{h}$ under all the constraints in (43) with an extra constraint $\text{tr}\left(\tilde{\mathbf{S}}_{11}\right) \leq \text{tr}\left(\tilde{\mathbf{S}}_{11}^*\right)$. Therefore, to prove (54), it suffices to prove that the rank of the optimal covariance matrix for the following optimization problem is no greater than 1:

$$\begin{aligned} \max \quad & \boldsymbol{h}^{\dagger}\tilde{\mathbf{S}}_{11}\boldsymbol{h} \\ \text{subject to} \quad & \boldsymbol{h}_j^{\dagger}\tilde{\mathbf{S}}_{11}\boldsymbol{h}_j \leq z_{mj}^2, \quad j = 1, \cdots, m-1 \\ & \text{tr}\left(\tilde{\mathbf{S}}_{11}\right) \leq \bar{P}, \quad \tilde{\mathbf{S}}_{11} \succeq \mathbf{0}, \end{aligned} \tag{55}$$

where

$$\bar{P} = \text{tr}\left(\tilde{\mathbf{S}}_{11}^*\right) \leq P. \tag{56}$$

The equivalence is due to the fact that the optimal $\tilde{\mathbf{S}}_{11}^{*}$ for problem (43) is also optimal for problem (55) and vice versa because of (56). Moreover, since $\tilde{\mathbf{S}}_{11}^{*}$ is also optimal for problem (55), the inequality constraint tr $\left(\tilde{\mathbf{S}}_{11}\right) \leq \bar{P}$ is active.

The Lagrangian of problem (55) is

$$L = -\boldsymbol{h}^{\dagger}\tilde{\mathbf{S}}_{11}\boldsymbol{h} + \sum_{j=1}^{m-1} \lambda_j \left(\boldsymbol{h}_j^{\dagger}\tilde{\mathbf{S}}_{11}\boldsymbol{h}_j - z_{mj}^2\right) + \lambda_m \left[\text{tr}\left(\tilde{\mathbf{S}}_{11}\right) - \bar{P}\right] - \text{tr}\left(\mathbf{W}\tilde{\mathbf{S}}_{11}\right). \tag{57}$$

On setting $\frac{\partial L}{\partial \tilde{\mathbf{S}}_{11}} = \mathbf{0}$, we have

$$\begin{aligned} \mathbf{W} &= -\boldsymbol{h}\boldsymbol{h}^{\dagger} + \sum_{j=1}^{m-1} \lambda_j \boldsymbol{h}_j \boldsymbol{h}_j^{\dagger} + \lambda_m \mathbf{I} \\ &= \mathbf{C} + \lambda_m \mathbf{I} \end{aligned} \tag{58}$$

where

$$\mathbf{C} = \mathbf{H} * \text{diag}\left[-1, \lambda_1, \cdots, \lambda_{m-1}\right] * \mathbf{H}^{\dagger} \tag{59}$$

and $\mathbf{H} = [\boldsymbol{h}, \boldsymbol{h}_1, \cdots, \boldsymbol{h}_{m-1}]$ is an $\bar{m} \times m$ matrix.

We then introduce the following lemma which is an extension of Sylvester's Law of Inertia.

*Lemma 6:* [29, Theorem 7] Let $\mathbf{H}$ be an $s \times t$ matrix and $\mathbf{A}$ be an $t \times t$ Hermitian matrix. Let $\pi(\cdot)$ and $\upsilon(\cdot)$ denote, respectively, the numbers of positive and negative eigenvalues of a matrix argument. Then we have

$$\pi\left(\mathbf{H}\mathbf{A}\mathbf{H}^{\dagger}\right) \leq \pi\left(\mathbf{A}\right), \quad \text{and} \quad \upsilon\left(\mathbf{H}\mathbf{A}\mathbf{H}^{\dagger}\right) \leq \upsilon\left(\mathbf{A}\right).$$

By Lemma 6 and the Karush-Kuhn-Tucker (KKT) conditions that require $\lambda_i \geq 0$, $i = 1, \cdots, m$, we have

$$\upsilon(\mathbf{C}) \leq 1. \tag{60}$$

Since $\mathbf{C}$ is an $\bar{m} \times \bar{m}$ Hermitian matrix, we can write its eigenvalue decomposition as

$$\mathbf{C} = \mathbf{Q}\text{diag}\left(\eta_1, \cdots, \eta_{\bar{m}}\right)\mathbf{Q}^{\dagger} \tag{61}$$

where $\mathbf{Q}^{\dagger}\mathbf{Q} = \mathbf{I}$, and $\eta_i$'s are the eigenvalues in ascending order. From (60), we have

$$\eta_1 \leq 0 \tag{62}$$

$$\text{and} \quad \eta_j \geq 0, \quad j = 2, \cdots, \bar{m}. \tag{63}$$

Since

$$\mathbf{W} = \mathbf{C} + \lambda_m \mathbf{I} = \mathbf{Q}\text{diag}\left[\lambda_m + \eta_1, \lambda_m + \eta_2, \cdots, \lambda_m + \eta_{\bar{m}}\right]\mathbf{Q}^\dagger,$$

$\lambda_m + \eta_i$ is an eigenvalue of $\mathbf{W}$, $i = 1, \cdots, \bar{m}$.

Since the optimal $\tilde{\mathbf{S}}_{11}^*$ for problem (55) satisfies $\text{tr}\left(\tilde{\mathbf{S}}_{11}^*\right) = \bar{P}$, from the KKT conditions we have

$$\lambda_m > 0. \tag{64}$$

Using (62)-(64) and noticing that the KKT conditions require $\mathbf{W} \succeq \mathbf{0}$, we have

$$\lambda_m + \eta_1 \geq 0 \tag{65}$$

$$\text{and} \quad \lambda_m + \eta_j > 0, \quad j = 2, \cdots, \bar{m}. \tag{66}$$

For the optimal $\tilde{\mathbf{S}}_{11}^*$, from the KKT conditions, we have

$$\begin{aligned}
\text{tr}\left(\mathbf{W}\tilde{\mathbf{S}}_{11}^*\right) &= \text{tr}\left(\left(\mathbf{C} + \lambda_m \mathbf{I}\right)\tilde{\mathbf{S}}_{11}^*\right) \\
&= \text{tr}\left(\mathbf{Q}\text{diag}\left[\lambda_m + \eta_1, \lambda_m + \eta_2, \cdots, \lambda_m + \eta_{\bar{m}}\right]\mathbf{Q}^\dagger\tilde{\mathbf{S}}_{11}^*\right) \\
&= \text{tr}\left(\text{diag}\left[\lambda_m + \eta_1, \lambda_m + \eta_2, \cdots, \lambda_m + \eta_{\bar{m}}\right]\mathbf{Q}^\dagger\tilde{\mathbf{S}}_{11}^*\mathbf{Q}\right) \\
&= \sum_{i=1}^{\bar{m}}(\lambda_m + \eta_i)\left(\mathbf{Q}^\dagger\tilde{\mathbf{S}}_{11}^*\mathbf{Q}\right)_{ii} \\
&= 0.
\end{aligned}$$

Since $\left(\mathbf{Q}^\dagger\tilde{\mathbf{S}}_{11}^*\mathbf{Q}\right)_{ii} \geq 0$, using (65) and (66) we have

$$\begin{aligned}
&\left(\mathbf{Q}^\dagger\tilde{\mathbf{S}}_{11}^*\mathbf{Q}\right)_{11} \geq 0, \\
&\left(\mathbf{Q}^\dagger\tilde{\mathbf{S}}_{11}^*\mathbf{Q}\right)_{jj} = 0, \quad j = 2, \cdots, \bar{m},
\end{aligned}$$

and $\left(\mathbf{Q}^\dagger\tilde{\mathbf{S}}_{11}^*\mathbf{Q}\right)_{11}$ can be non-zero if $\lambda_m + \eta_1 = 0$. Since $\mathbf{Q}^\dagger\tilde{\mathbf{S}}_{11}^*\mathbf{Q} \succeq \mathbf{0}$, if one diagonal element is zero then all the elements on this row and this column must be zero. Thus, we have

$$\text{rank}\left(\mathbf{S}_m^*\right) \leq \text{rank}\left(\tilde{\mathbf{S}}_{11}^*\right) = \text{rank}\left(\mathbf{Q}^\dagger\tilde{\mathbf{S}}_{11}^*\mathbf{Q}\right) \leq 1.$$

■

Theorem 3 proves the sufficiency of transmitter beamforming for achieving the SUD rate region. However, it does not mean that the SUD rate region can be achieved only by beamforming. As shown in the examples, the optimal $\mathbf{S}_m^*$ is not unique when $\|\boldsymbol{h}\| = 0$ or $\left\|\widehat{\boldsymbol{h}}\right\| = 0$. $\|\boldsymbol{h}\| = 0$ corresponds to the case in which $\boldsymbol{h}_{mm}$ is orthogonal to $\boldsymbol{h}_{mj}$, $j = 1, \cdots, m-1$, and $\left\|\widehat{\boldsymbol{h}}\right\| = 0$ corresponds to the case in which $\boldsymbol{h}_{mm}$ is linearly dependent of $\boldsymbol{h}_{mj}$, $j = 1, \cdots, m-1$.

### B. SUD rate region of $m$-user MISO ICs

Section III-A establishes the optimality of beamforming for the general $m$-user MISO IC with complex channels when each receiver is restricted to SUD. In this section, we first use a three-user real MISO IC as an example to show how we can obtain the SUD rate region by Theorem 3, and then generalize it to $m$-user complex MISO ICs.

For a three-user MISO IC, the optimization problem (42) can be written as

$$\begin{aligned} \max \quad & \boldsymbol{h}_0^T \mathbf{S} \boldsymbol{h}_0 \\ \text{subject to} \quad & \boldsymbol{h}_1^T \mathbf{S} \boldsymbol{h}_1 \leq z_1^2 \\ & \boldsymbol{h}_2^T \mathbf{S} \boldsymbol{h}_2 \leq z_2^2 \\ & \text{tr}(\mathbf{S}) \leq P, \quad \mathbf{S} \succeq \mathbf{0}. \end{aligned} \tag{67}$$

We first reformulate this problem into (43). Let the SVD of $\boldsymbol{h}_1$ be

$$\boldsymbol{h}_1 = \mathbf{U}_1 \begin{bmatrix} \|\boldsymbol{h}_1\| \\ \mathbf{0} \end{bmatrix},$$

where $\mathbf{U}_1^T \mathbf{U}_1 = \mathbf{I}$. We then update $\boldsymbol{h}_0$, $\boldsymbol{h}_1$ $\boldsymbol{h}_2$ and $\mathbf{S}$ as follows:

$$\boldsymbol{h}_0^{(1)} = \mathbf{U}_1^T \boldsymbol{h}_0, \tag{68}$$

$$\boldsymbol{h}_1^{(1)} = \mathbf{U}_1^T \boldsymbol{h}_1 = \begin{bmatrix} \|\boldsymbol{h}_1\| \\ \mathbf{0} \end{bmatrix}, \tag{69}$$

$$\boldsymbol{h}_2^{(1)} = \mathbf{U}_1^T \boldsymbol{h}_2 \triangleq \begin{bmatrix} \widehat{h}_{21} \\ \widehat{\boldsymbol{h}}_{22} \end{bmatrix} \tag{70}$$

$$\text{and} \quad \mathbf{S}^{(1)} = \mathbf{U}_1^T \mathbf{S}_1 \mathbf{U}_1, \tag{71}$$

where $\widehat{h}_{21} = \left(\boldsymbol{h}_2^{(1)}\right)_1$, and $\widehat{\boldsymbol{h}}_{22}$ is the remaining part of $\widehat{\boldsymbol{h}}_2^{(1)}$. Let the SVD of $\widehat{\boldsymbol{h}}_{22}$ be

$$\widehat{\boldsymbol{h}}_{22} = \mathbf{U}_2 \begin{bmatrix} \left\|\widehat{\boldsymbol{h}}_{22}\right\| \\ \mathbf{0} \end{bmatrix},$$

and $\mathbf{U}_2^T\mathbf{U}_2 = \mathbf{I}$. Then we update $\boldsymbol{h}_0$, $\boldsymbol{h}_1$ $\boldsymbol{h}_2$ and $\mathbf{S}$ as follows:

$$\boldsymbol{h}_0^{(2)} = \begin{bmatrix} 1 & \mathbf{0} \\ \mathbf{0} & \mathbf{U}_2 \end{bmatrix}^T \mathbf{U}_1^T \boldsymbol{h}_0 \triangleq \begin{bmatrix} \tilde{\boldsymbol{h}}_{01} \\ \tilde{\boldsymbol{h}}_{02} \end{bmatrix},$$

$$\boldsymbol{h}_1^{(2)} = \begin{bmatrix} 1 & \mathbf{0} \\ \mathbf{0} & \mathbf{U}_2 \end{bmatrix}^T \begin{bmatrix} \|\boldsymbol{h}_1\| \\ \mathbf{0} \end{bmatrix} = \begin{bmatrix} \|\boldsymbol{h}_1\| \\ \mathbf{0} \end{bmatrix},$$

$$\boldsymbol{h}_2^{(2)} = \begin{bmatrix} 1 & \mathbf{0} \\ \mathbf{0} & \mathbf{U}_2 \end{bmatrix}^T \begin{bmatrix} \widehat{h}_{21} \\ \widehat{\boldsymbol{h}}_{22} \end{bmatrix} = \begin{bmatrix} \widehat{h}_{21} \\ \left\|\widehat{\boldsymbol{h}}_{22}\right\| \\ \mathbf{0} \end{bmatrix}$$

$$\text{and} \quad \mathbf{S}^{(2)} = \begin{bmatrix} 1 & \mathbf{0} \\ \mathbf{0} & \mathbf{U}_2 \end{bmatrix}^T \mathbf{U}_1^T \mathbf{S}_1 \mathbf{U}_1 \begin{bmatrix} 1 & \mathbf{0} \\ \mathbf{0} & \mathbf{U}_2 \end{bmatrix},$$

where $\tilde{\boldsymbol{h}}_{11}$ is a $2 \times 1$ vector. Define

$$\tilde{\mathbf{S}} = \mathbf{S}^{(2)} = \begin{bmatrix} \widetilde{\mathbf{S}}_{11} & \widetilde{\mathbf{S}}_{21}^T \\ \widetilde{\mathbf{S}}_{21} & \widetilde{\mathbf{S}}_{22} \end{bmatrix} \tag{72}$$

where $\widetilde{\mathbf{S}}_{11}$ is a $2 \times 2$ matrix. Then we obtain the following optimization problem:

$$\begin{aligned} \max \quad & \left( \sqrt{\tilde{\boldsymbol{h}}_{01}^T \tilde{\mathbf{S}}_{11} \tilde{\boldsymbol{h}}_{01}} + \|\tilde{\boldsymbol{h}}_{02}\| \sqrt{P - \text{tr}\left(\tilde{\mathbf{S}}_{11}\right)} \right)^2 \\ \text{subject to} \quad & \begin{bmatrix} \|\boldsymbol{h}_1\| \\ 0 \end{bmatrix}^T \tilde{\mathbf{S}}_{11} \begin{bmatrix} \|\boldsymbol{h}_1\| \\ 0 \end{bmatrix} \le z_1^2 \\ & \begin{bmatrix} \widehat{h}_{21} \\ \left\|\widehat{h}_{22}\right\| \end{bmatrix}^T \tilde{\mathbf{S}}_{11} \begin{bmatrix} \widehat{h}_{21} \\ \left\|\widehat{h}_{22}\right\| \end{bmatrix} \le z_2^2 \\ & \text{tr}\left(\tilde{\mathbf{S}}_{11}\right) \le P, \quad \tilde{\mathbf{S}}_{11} \succeq \mathbf{0}. \end{aligned} \tag{73}$$

The solution for problem (73) is complex. In the following we obtain the SUD rate region by using the fact that the rank of the optimal $\tilde{\mathbf{S}}_{11}$ is no greater than one without directly solving problem (73). Therefore, instead of exhausting over all feasible $z_1^2$ and $z_2^2$ and collect all corresponding $\tilde{\mathbf{S}}_{11}$, we exhaust over all feasible $\tilde{\mathbf{S}}_{11}$. We let

$$\widetilde{\mathbf{S}}_{11} = P \begin{bmatrix} \sin^2 \psi_1 & \cos\psi_1 \sin\psi_1 \sin\psi_2 \\ \cos\psi_1 \sin\psi_1 \sin\psi_2 & \cos^2\psi_1 \sin^2\psi_2 \end{bmatrix}, \tag{74}$$

where $\psi_1$ and $\psi_2$ can be any values in $[0,\pi]$. Then we have that the optimal $\mathbf{S}$ for this $\tilde{\mathbf{S}}_{11}$ is

$$\mathbf{S}\left(\boldsymbol{h}_0,\boldsymbol{h}_1,\boldsymbol{h}_2,P,\psi_1,\psi_2\right)$$
$$=\begin{cases}\mathbf{U}_1\begin{bmatrix}1 & \mathbf{0}\\ \mathbf{0} & \mathbf{U}_2\end{bmatrix}\begin{bmatrix}\tilde{\mathbf{S}}_{11} & \frac{\sqrt{P-\mathrm{tr}\left(\tilde{\mathbf{S}}_{11}\right)}}{\left\|\tilde{\boldsymbol{h}}_{02}\right\|\sqrt{\tilde{\boldsymbol{h}}_{01}^T\tilde{\mathbf{S}}_{11}\tilde{\boldsymbol{h}}_{01}}}\tilde{\mathbf{S}}_{11}\tilde{\boldsymbol{h}}_{01}\tilde{\boldsymbol{h}}_{02}^T\\ \frac{\sqrt{P-\mathrm{tr}\left(\tilde{\mathbf{S}}_{11}\right)}}{\left\|\tilde{\boldsymbol{h}}_{02}\right\|\sqrt{\tilde{\boldsymbol{h}}_{01}^T\tilde{\mathbf{S}}_{11}\tilde{\boldsymbol{h}}_{01}}}\tilde{\boldsymbol{h}}_{02}\tilde{\boldsymbol{h}}_{01}^T\tilde{\mathbf{S}}_{11} & \frac{P-\mathrm{tr}\left(\tilde{\mathbf{S}}_{11}\right)}{\left\|\tilde{\boldsymbol{h}}_{02}\right\|^2}\tilde{\boldsymbol{h}}_{02}\tilde{\boldsymbol{h}}_{02}^T\end{bmatrix}\begin{bmatrix}1 & \mathbf{0}\\ \mathbf{0} & \mathbf{U}_2^T\end{bmatrix}\mathbf{U}_1^T\\ \qquad\qquad\text{when } \tilde{\boldsymbol{h}}_{01}^T\tilde{\mathbf{S}}_{11}\tilde{\boldsymbol{h}}_{01}\neq 0,\ \left\|\tilde{\boldsymbol{h}}_{02}\right\|\neq 0\\ \mathbf{U}_1\begin{bmatrix}1 & \mathbf{0}\\ \mathbf{0} & \mathbf{U}_2\end{bmatrix}\begin{bmatrix}\tilde{\mathbf{S}}_{11} & \frac{\sqrt{P-\mathrm{tr}\left(\tilde{\mathbf{S}}_{11}\right)}}{\left\|\tilde{\boldsymbol{h}}_{02}\right\|}\tilde{\mathbf{S}}_{11}^{\frac{T}{2}}\mathbf{1}_0\tilde{\boldsymbol{h}}_{02}^T\\ \frac{\sqrt{P-\mathrm{tr}\left(\tilde{\mathbf{S}}_{11}\right)}}{\left\|\tilde{\boldsymbol{h}}_{02}\right\|}\tilde{\boldsymbol{h}}_{02}\mathbf{1}_0^T\tilde{\mathbf{S}}_{11}^{\frac{1}{2}} & \frac{P-\mathrm{tr}\left(\tilde{\mathbf{S}}_{11}\right)}{\left\|\tilde{\boldsymbol{h}}_{02}\right\|^2}\tilde{\boldsymbol{h}}_{02}\tilde{\boldsymbol{h}}_{02}^T\end{bmatrix}\begin{bmatrix}1 & \mathbf{0}\\ \mathbf{0} & \mathbf{U}_2^T\end{bmatrix}\mathbf{U}_1^T\\ \qquad\qquad\text{when } \tilde{\boldsymbol{h}}_{01}^T\tilde{\mathbf{S}}_{11}\tilde{\boldsymbol{h}}_{01}=0,\ \left\|\tilde{\boldsymbol{h}}_{02}\right\|\neq 0\\ \mathbf{U}_1\begin{bmatrix}1 & \mathbf{0}\\ \mathbf{0} & \mathbf{U}_2\end{bmatrix}\begin{bmatrix}\tilde{\mathbf{S}}_{11} & \mathbf{0}\\ \mathbf{0} & \mathbf{0}\end{bmatrix}\begin{bmatrix}1 & \mathbf{0}\\ \mathbf{0} & \mathbf{U}_2^T\end{bmatrix}\mathbf{U}_1^T \qquad \text{when } \left\|\tilde{\boldsymbol{h}}_{02}\right\|=0.\end{cases}$$

It can be shown that

$$\begin{bmatrix}\widehat{h}_{21}\\ \left\|\widehat{\boldsymbol{h}}_{22}\right\|\end{bmatrix}=\|\boldsymbol{h}_2\|\begin{bmatrix}\cos\theta_{12}\\ \sin\theta_{12}\end{bmatrix}$$
$$\widetilde{\boldsymbol{h}}_{01}=\|\boldsymbol{h}_0\|\begin{bmatrix}\cos\theta_{01}\\ \sin\theta_{01}\cos\widehat{\theta}\end{bmatrix},$$

where

$$\theta_{ij}=\angle\left(\boldsymbol{h}_i,\boldsymbol{h}_j\right),\quad i,j=0,1,2,$$
$$\text{and}\quad \widehat{\theta}=\angle\left(\widehat{\boldsymbol{h}}_{02},\widehat{\boldsymbol{h}}_{22}\right)=\begin{cases}\cos^{-1}\left(\frac{\cos\theta_{02}-\cos\theta_{01}\cos\theta_{12}}{\sin\theta_{01}\sin\theta_{12}}\right), & \theta_{01}\neq 0,\pi,\quad \theta_{12}\neq 0,\pi\\ \pi/2, & \text{otherwise.}\end{cases}$$

$\widehat{\theta}$ is the angle between the projections of $\boldsymbol{h}_0$ and $\boldsymbol{h}_2$ in $\boldsymbol{h}_1$'s orthogonal subspace. The signal power and the associated interference powers are determined by $\tilde{\mathbf{S}}_{11}$, and from (73) we have

$$\boldsymbol{h}_0^T\mathbf{S}\boldsymbol{h}_0=P\left\|\boldsymbol{h}_0\right\|^2\left(\left|\cos\theta_{01}\sin\psi_1+\sin\theta_{01}\cos\widehat{\theta}\cos\psi_1\sin\psi_2\right|+\sin\theta_{01}\sin\widehat{\theta}\left|\cos\psi_1\cos\psi_2\right|\right)^2 \tag{75}$$
$$\boldsymbol{h}_1^T\mathbf{S}\boldsymbol{h}_1=P\left\|\boldsymbol{h}_1\right\|^2\sin^2\psi_1 \tag{76}$$
$$\text{and}\quad \boldsymbol{h}_2^T\mathbf{S}\boldsymbol{h}_2=P\left\|\boldsymbol{h}_2\right\|^2\left(\cos\theta_{12}\sin\psi_1+\sin\theta_{12}\cos\psi_1\sin\psi_2\right)^2. \tag{77}$$

Therefore, we have the following theorem.

*Theorem 4:* The SUD rate region of a three-user MISO IC is

$$\bigcup_{\substack{\psi_i\in[0,\pi]\\ i=1,\cdots,6}} \left\{ \begin{array}{l} R_1 \le \dfrac{1}{2}\log\left[1+\dfrac{\boldsymbol{h}_{11}^T\mathbf{S}(\boldsymbol{h}_{11},\boldsymbol{h}_{12},\boldsymbol{h}_{13},P_1,\psi_1,\psi_2)\boldsymbol{h}_{11}}{1+\boldsymbol{h}_{21}^T\mathbf{S}(\boldsymbol{h}_{22},\boldsymbol{h}_{21},\boldsymbol{h}_{23},P_2,\psi_3,\psi_4)\boldsymbol{h}_{21}+\boldsymbol{h}_{31}^T\mathbf{S}(\boldsymbol{h}_{33},\boldsymbol{h}_{31},\boldsymbol{h}_{32},P_3,\psi_5,\psi_6)\boldsymbol{h}_{31}}\right] \\ R_2 \le \dfrac{1}{2}\log\left[1+\dfrac{\boldsymbol{h}_{22}^T\mathbf{S}(\boldsymbol{h}_{22},\boldsymbol{h}_{21},\boldsymbol{h}_{23},P_2,\psi_3,\psi_4)\boldsymbol{h}_{22}}{1+\boldsymbol{h}_{12}^T\mathbf{S}(\boldsymbol{h}_{11},\boldsymbol{h}_{12},\boldsymbol{h}_{13},P_1,\psi_1,\psi_2)\boldsymbol{h}_{12}+\boldsymbol{h}_{32}^T\mathbf{S}(\boldsymbol{h}_{33},\boldsymbol{h}_{31},\boldsymbol{h}_{32},P_3,\psi_5,\psi_6)\boldsymbol{h}_{32}}\right] \\ R_3 \le \dfrac{1}{2}\log\left[1+\dfrac{\boldsymbol{h}_{33}^T\mathbf{S}(\boldsymbol{h}_{33},\boldsymbol{h}_{31},\boldsymbol{h}_{32},P_3,\psi_5,\psi_6)\boldsymbol{h}_{33}}{1+\boldsymbol{h}_{13}^T\mathbf{S}(\boldsymbol{h}_{11},\boldsymbol{h}_{12},\boldsymbol{h}_{13},P_1,\psi_1,\psi_2)\boldsymbol{h}_{13}+\boldsymbol{h}_{23}^T\mathbf{S}(\boldsymbol{h}_{22},\boldsymbol{h}_{21},\boldsymbol{h}_{23},P_2,\psi_3,\psi_4)\boldsymbol{h}_{23}}\right] \end{array} \right\} \tag{78}$$

Furthermore, the boundary points of the rate region can be achieved by restricting each transmitter to implement beamforming.

In the following, we discuss some special rate triples of the above region. For simplicity, we assume $0 \le \angle(\boldsymbol{h}_{ij}, \boldsymbol{h}_{ik}) \le \frac{\pi}{2}$, where $i, j$, and $k$ are integers ranging from $1$ to 3.

- ZF beamforming rate triple.

  When $\psi_1 = \psi_2 = 0$, (75)-(77) become

  $$\begin{aligned} &\boldsymbol{h}_0^T\mathbf{S}\boldsymbol{h}_0 = P\|\boldsymbol{h}_0\|^2\sin^2\theta_{01}\sin^2\widehat{\theta}, \\ &\boldsymbol{h}_1^T\mathbf{S}\boldsymbol{h}_1 = 0 \\ \text{and}\quad &\boldsymbol{h}_2^T\mathbf{S}\boldsymbol{h}_2 = 0. \end{aligned}$$

  Therefore, the rate triple in set (78) with $\psi_i = 0$, $i = 1, \cdots, 6$ is the ZF beamforming rate triple. That is, the beamforming vector is chosen to be orthogonal to the channel vectors corresponding to the interference links associated with the same transmitter.

- Single-user maximum rate surface.

  When $\psi_1 = \frac{\pi}{2} - \theta_{01}$ and $\psi_2 = \frac{\pi}{2} - \widehat{\theta}$, (75)-(77) become

  $$\begin{aligned} &\boldsymbol{h}_0^T\mathbf{S}\boldsymbol{h}_0 = P\|\boldsymbol{h}_0\|^2, \\ &\boldsymbol{h}_1^T\mathbf{S}\boldsymbol{h}_1 = P\|\boldsymbol{h}_1\|^2\cos^2\theta_{01} \\ \text{and}\quad &\boldsymbol{h}_2^T\mathbf{S}\boldsymbol{h}_2 = P\|\boldsymbol{h}_2\|^2\cos^2\theta_{02}. \end{aligned}$$

  Therefore, $\boldsymbol{h}_0^T\mathbf{S}\boldsymbol{h}_0$ achieves the maximum with the constraint $\text{tr}(\mathbf{S}) \le P$. Then we obtain the following result. Let

  $$\psi_1 = \frac{\pi}{2} - \angle(\boldsymbol{h}_{11}, \boldsymbol{h}_{12}), \quad \psi_2 = \frac{\pi}{2} - \cos^{-1}\left(\frac{\cos\angle(\boldsymbol{h}_{11},\boldsymbol{h}_{13}) - \cos\angle(\boldsymbol{h}_{11},\boldsymbol{h}_{12})\cos\angle(\boldsymbol{h}_{12},\boldsymbol{h}_{13})}{\sin\angle(\boldsymbol{h}_{11},\boldsymbol{h}_{12})\sin\angle(\boldsymbol{h}_{12},\boldsymbol{h}_{13})}\right),$$
  $$\psi_3 = \psi_5 = 0, \qquad \psi_4 \in [0,\pi] \quad \text{and} \quad \psi_6 \in [0,\pi];$$

then the corresponding rate triples form the surface of the three-dimensional (3-D) SUD rate region with $R_1$ at its maximum. Similarly, the maximum rate surface for users 2 and 3 can be obtained.

- Projection of the 3-D rate region to the two-dimensional (2-D) rate region.

  Suppose, say, user 1 is silent. We can recover the 2-D SUD rate region formed by users 2 and 3 obtained from Theorem 1. That is, the surface of the 3-D SUD rate region corresponding to $R_1 = 0$ is precisely the 2-D SUD rate region of the two-user IC consisting of users 2 and 3.

We can similarly obtain the SUD rate region of an $m$-user complex MISO IC with $m \geq 3$ by Theorem 3. The only difference is how we generate matrix $\tilde{\mathbf{S}}_{11}$. We generalize it as follows:

$$\tilde{\mathbf{S}}_{11} = P\tilde{\boldsymbol{\gamma}}\tilde{\boldsymbol{\gamma}}^{\dagger},$$

$$(\tilde{\boldsymbol{\gamma}})_1 = e^{j\omega_1} \sin\psi_1$$

$$\text{and} \quad (\tilde{\boldsymbol{\gamma}})_i = e^{j\omega_i} \sin\psi_i \sqrt{1 - \sum_{k=1}^{i-1} |\tilde{\boldsymbol{\gamma}}|_k^2}, \quad i = 2, \cdots \bar{m},$$

where $\psi_i \in [0, \pi]$, $\omega_i \in [0, 2\pi)$, and $j$ denotes the imaginary unit. In the case of a real MISO IC, we can simply choose $\omega_i = 0$. The rate region and the corresponding beamforming vectors can be obtained in the same way as in Theorem 4, and so the determination of their quantities is omitted.

## IV. Numerical examples

In this section, we present the SUD rate region of two-user or three-user real MISO ICs. For ease of visualization, we take the convex hull of all the regions.

The achievable rate regions for the symmetric MISO ICs are shown in Figs. 3 and 4. Here, a symmetric MISO IC refers to one with $\|\boldsymbol{h}_1\| = \|\boldsymbol{h}_4\|$, $\|\boldsymbol{h}_2\| = \|\boldsymbol{h}_3\|$, $\theta_1 = \theta_2$, and $P_1 = P_2$. In Fig. 3, the rate region shrinks as $\theta$ varies from $\frac{\pi}{2}$ to $0$, which corresponds to the MISO IC with the interference link and the direct link varying from being orthogonal to co-linear. The SUD rate region becomes smaller than that of frequency division multiplexing (FDM) when the direct link and interference link exhibit increasing linear dependence. When $\theta = \frac{\pi}{2}$, neither of the transmitters generates interference to the other user. Therefore, the IC reduces to two parallel single-user channels without interference, and the ZF beamforming rates achieve the maximum. When $\theta = 0$, the two transmitters generate the worst interference and this MISO IC acts as a single antenna IC. Hence, the ZF beamforming rates are $0$.

In Fig. 4, the rate region also shrinks as the interference gain $\sigma$ increases. However, even if $\sigma \to \infty$ the rate region will not be worse than the tetragon $OMAN$. Points $M, N$ are the extreme points of the rate region on the axes. Point $A$ denotes the ZF beamforming rates, which are determined only by $P_1, P_2$ and $\theta$. Therefore, as $\sigma$ increases, the rate region becomes close to the tetragon. Point $A$ also shows the advantage of multiple antenna systems over single antenna systems, since the achievable rate region for the latter case reduces to the triangle defined by $OMN$ when $\sigma \to \infty$. It can be shown that point $A$ falls inside the FDM region only if

$$\sin\theta \leq \sqrt{\frac{\sqrt{1+2P}-1}{P}}, \tag{79}$$

i.e., FDM outperforms SUD when the interference link subspace is close to the direct link subspace and interference gain is sufficiently large. Since

$$\lim_{P\to\infty}\sqrt{\frac{\sqrt{1+2P}-1}{P}} = 0, \text{ and } \lim_{P\to 0}\sqrt{\frac{\sqrt{1+2P}-1}{P}} = 1,$$

for all MISO ICs with $\theta_1 \neq 0, \frac{\pi}{2}$ and $\theta_2 \neq 0, \frac{\pi}{2}$, SUD achieves larger rate region than FDM for large power constraints, while the reverse is true for small power constraints and sufficiently large interference gains $\sigma^2$.

Fig. 5 shows the SUD rate region of a two-user MISO IC under an interference power constraint. When such a constraint is small, this corresponding rate region is included in the FDM rate region. Since neither user generates interference in FDM, when the interference power is a concern in the system design, one can choose FDM instead of SUD when the interference is restricted.

Fig. 6 shows the SUD rate region of a three-user MISO IC with the power constraints $P_1 = 1$, $P_2 = 1.5$ and $P_3 = 2$. The channels are

$$\mathbf{H}_1 = \begin{bmatrix} -2.1 & 0 & 0.5 \\ 0.1 & 0.2 & 0.1 \\ 1.5 & 0.9 & 0.3 \\ 0.1 & 0.2 & -1 \\ 0.2 & 0.8 & -0.9 \end{bmatrix}, \quad \mathbf{H}_2 = \begin{bmatrix} 0 & 2.7 & -0.5 \\ 0.4 & 0.4 & 0.2 \\ -0.9 & -1.3 & -0.6 \\ 0.8 & 0.4 & 0 \\ 0.1 & 0.5 & 0.4 \end{bmatrix} \text{ and } \mathbf{H}_3 = \begin{bmatrix} 1.2 & 0 & 1 \\ 0.8 & 0.9 & -1.7 \\ -2.6 & 0.8 & -1 \\ 0.3 & 1.3 & 0.7 \\ 0.8 & 1.2 & -1 \end{bmatrix},$$

where $\mathbf{H}_1 = [\boldsymbol{h}_{11}, \boldsymbol{h}_{12}, \boldsymbol{h}_{13}]$, $\mathbf{H}_2 = [\boldsymbol{h}_{21}, \boldsymbol{h}_{22}, \boldsymbol{h}_{23}]$ and $\mathbf{H}_3 = [\boldsymbol{h}_{31}, \boldsymbol{h}_{32}, \boldsymbol{h}_{33}]$. The solid curves are the rate regions for one user being inactive or at the maximum rate. That is, they are the projection of the 3-D rate region onto a 2-D plane with one rate fixed at a constant value. The ZF beamforming rate triple of this channel is shown in Fig. 7. The two cutting planes that pass through this rate point show that the ZF beamforming rate point is not on the boundary of the SUD rate region.

## V. Conclusion

We have considered MISO ICs where each transmitter is limited to a Gaussian input and each receiver is limited to single-user detection. By exploiting the relation between the signal power at the intended receiver and the interference power at the unintended receiver, we have derive a new method to obtain the SUD rate region for the MISO IC. We have shown that the original family of non-convex optimization problems is readily reduced to an equivalent family of convex optimization problems whose closed-form solutions are provided. As a consequence of restricting each receiver to implement single-user detection, transmitter beamforming is shown to be sufficient to achieve all boundary points of the SUD rate region.

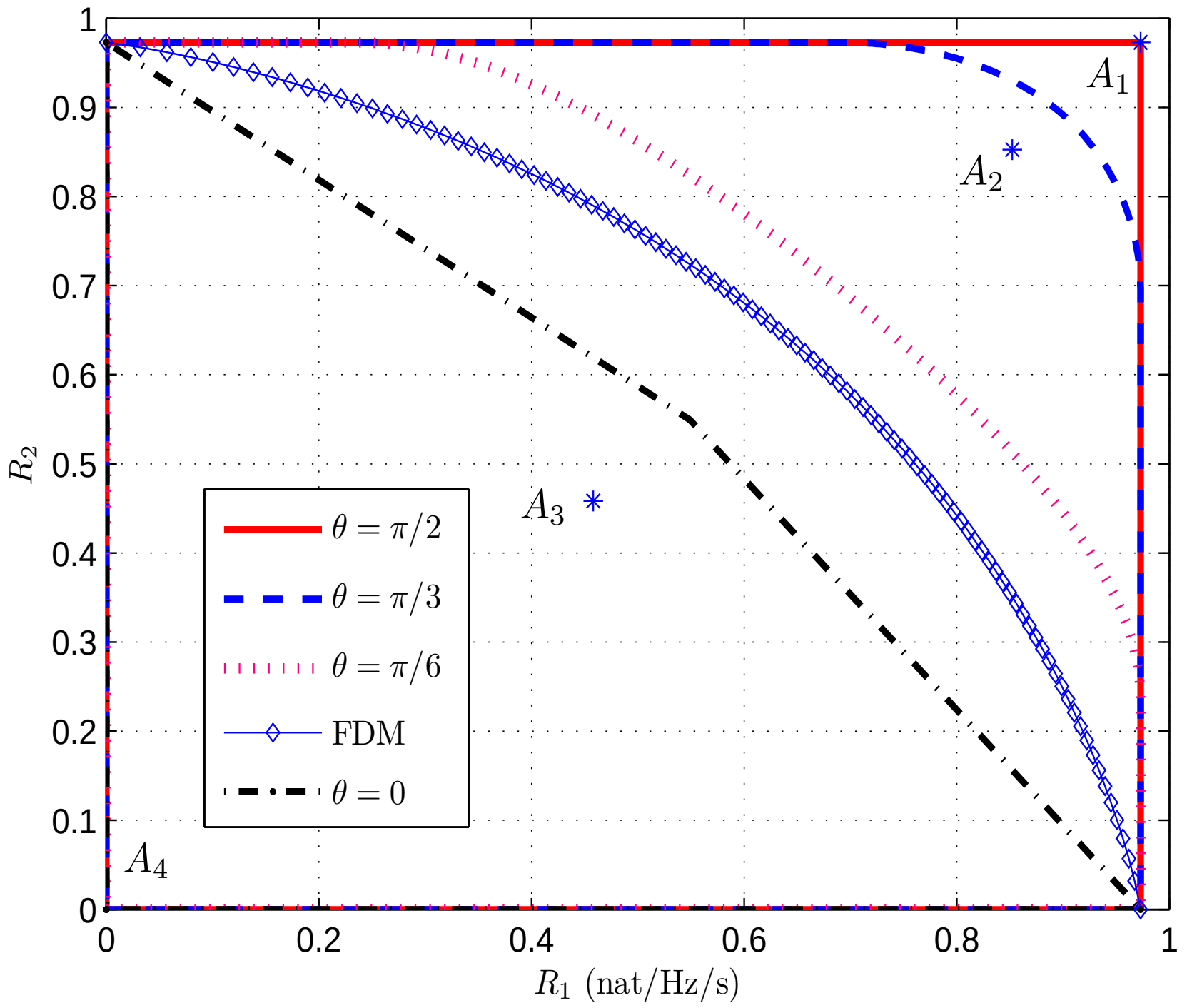

Fig. 3. The achievable regions for the symmetric MISO Gaussian IC with $\|\boldsymbol{h}_1\| = \|\boldsymbol{h}_4\| = 1$, $\|\boldsymbol{h}_2\| = \|\boldsymbol{h}_3\| = \frac{1}{\sqrt{3}}$, $\theta_1 = \theta_2 = \theta$, and $P_1 = P_2 = 6$. $A_1$-$A_4$ denote the respective ZF beamforming rates for $\theta = \frac{\pi}{2}$ to $\theta = 0$. Also plotted for comparison is the FDM achievable rate region.

# Appendix

## A. Proof of Lemma 2

Since it is straightforward to show (16)-(19) when $\boldsymbol{h}_1$ and $\boldsymbol{h}_3$ are linearly dependent, we only need to prove (12)-(15) under the condition that $\boldsymbol{h}_1$ and $\boldsymbol{h}_3$ are linearly independent.

Define

$$\widehat{\mathbf{S}}_1 = \mathbf{U}_3^{\dagger}\mathbf{S}_1\mathbf{U}_3 = \begin{bmatrix} \widehat{S}_{11} & \boldsymbol{\alpha}^{\dagger} \\ \boldsymbol{\alpha} & \mathbf{A} \end{bmatrix}, \tag{80}$$

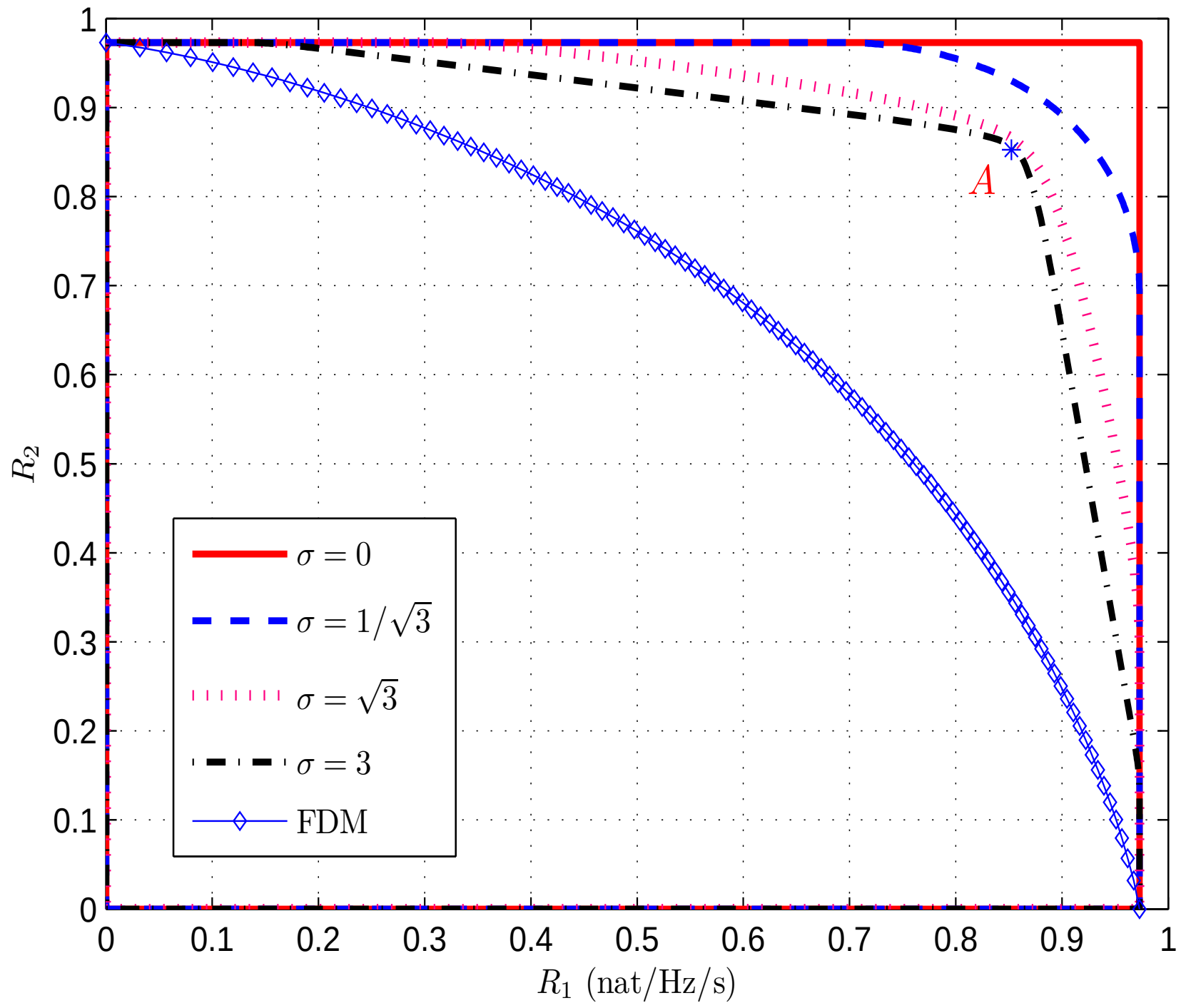

Fig. 4. The achievable regions for the symmetric MISO Gaussian IC with $\|\boldsymbol{h}_1\| = \|\boldsymbol{h}_4\| = 1, \|\boldsymbol{h}_2\| = \|\boldsymbol{h}_3\| = \sigma, \theta_1 = \theta_2 = \frac{\pi}{3}$, and $P_1 = P_2 = 6$. $A$ is the ZF beamforming rate point for all the choices of $\sigma$. Also plotted for comparison is the FDM achievable rate region.

where $\widehat{S}_{11} = \left(\widehat{\mathbf{S}}_1\right)_{11}$ is a non-negative real scalar, $\boldsymbol{\alpha}$ is a column vector, and $\mathbf{A}$ is a Hermitian matrix. Since $\widehat{\mathbf{S}}_1 \succeq \mathbf{0}$, we have

$$\widehat{S}_{11}\mathbf{A} \succeq \boldsymbol{\alpha}\boldsymbol{\alpha}^\dagger. \tag{81}$$

On Substituting (80) and (11) into (4), we have

$$\boldsymbol{h}_1^\dagger \mathbf{S}_1 \boldsymbol{h}_1 = \left|\widehat{h}_{11}\right|^2 \widehat{S}_{11} + \widehat{h}_{11}\boldsymbol{\beta}^\dagger\boldsymbol{\alpha} + \widehat{h}_{11}^\dagger \boldsymbol{\alpha}^\dagger\boldsymbol{\beta} + \boldsymbol{\beta}^\dagger\mathbf{A}\boldsymbol{\beta} \tag{82}$$

$$\text{and} \quad \boldsymbol{h}_3^\dagger \mathbf{S}_1 \boldsymbol{h}_3 = \|\boldsymbol{h}_3\|^2 \widehat{S}_{11} = z_1^2, \tag{83}$$

where $\|\boldsymbol{h}_3\| \neq 0$ since $\boldsymbol{h}_3$ is linearly independent of $\boldsymbol{h}_1$. Therefore the optimization problem (4) is

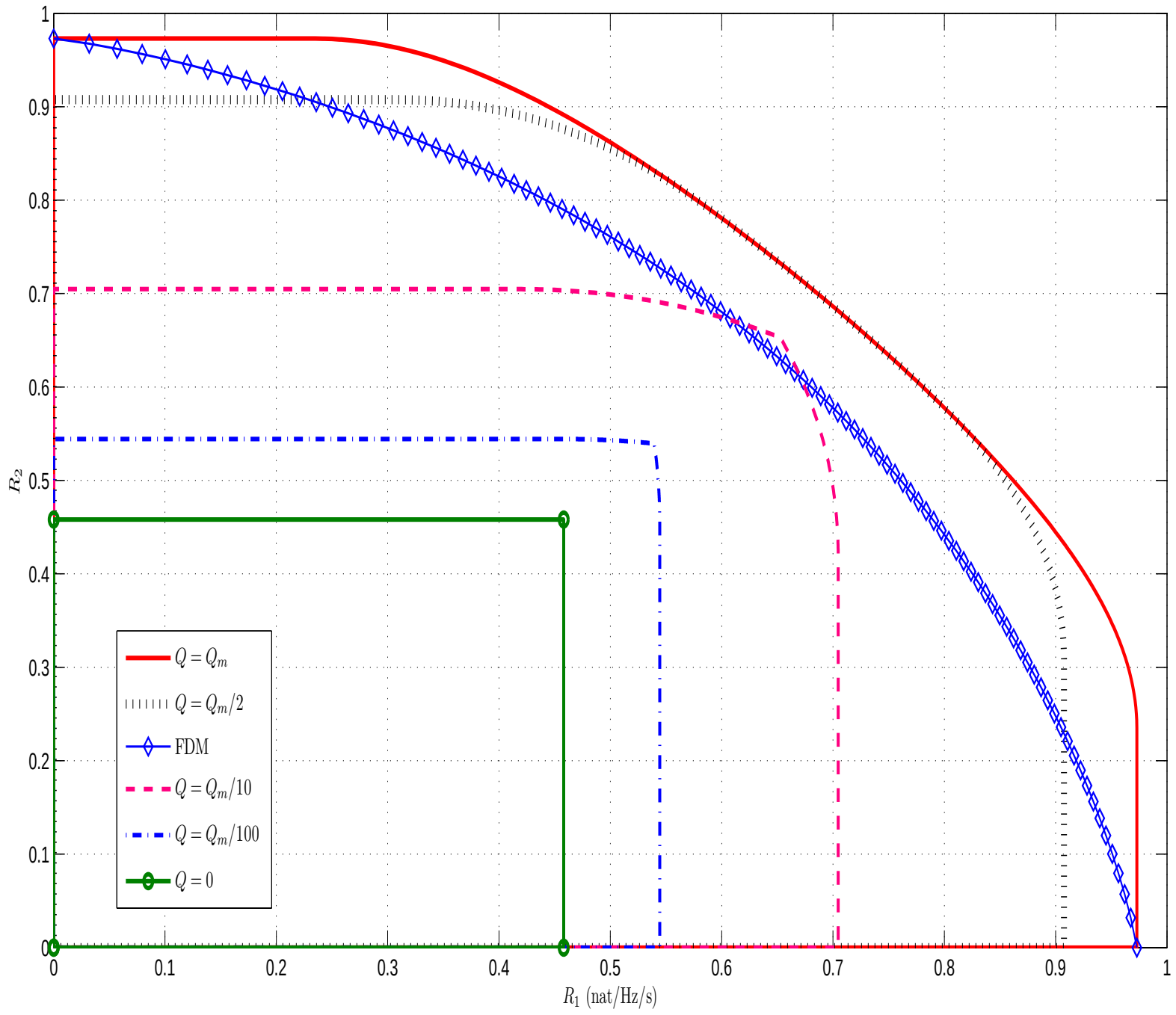

Fig. 5. The interference-limited SUD rate regions of a two-user MISO IC, where $\|\boldsymbol{h}_1\| = \|\boldsymbol{h}_4\| = 1$, $\|\boldsymbol{h}_2\| = \|\boldsymbol{h}_3\| = \frac{1}{\sqrt{3}}$, $\theta_1 = \theta_2 = \frac{\pi}{3}$, $P_1 = P_2 = 6$, and $Q_{\max} = P_1 \|\boldsymbol{h}_3\|^2 \cos^2\left(\frac{\pi}{3}\right)$.

equivalent to

$$\max \quad \frac{z_1^2}{\|\boldsymbol{h}_3\|^2}\left|\widehat{h}_{11}\right|^2 + \widehat{h}_{11}\boldsymbol{\beta}^\dagger\boldsymbol{\alpha} + \widehat{h}_{11}^\dagger\boldsymbol{\alpha}^\dagger\boldsymbol{\beta} + \boldsymbol{\beta}^\dagger\mathbf{A}\boldsymbol{\beta}$$

$$\text{subject to} \quad \frac{z_1^2}{\|\boldsymbol{h}_3\|^2}\mathbf{A} \succeq \boldsymbol{\alpha}\boldsymbol{\alpha}^\dagger, \tag{84}$$

$$\text{tr}(\mathbf{A}) \leq P_1 - \frac{z_1^2}{\|\boldsymbol{h}_3\|^2}. \tag{85}$$

Let the SVD of $\boldsymbol{\beta}$ be

$$\boldsymbol{\beta} = \mathbf{U}_\beta \begin{bmatrix} \|\boldsymbol{\beta}\| \\ \mathbf{0} \end{bmatrix}, \tag{86}$$

where $\mathbf{U}_\beta\mathbf{U}_\beta^\dagger = \mathbf{I}$. We further define

$$\widehat{\mathbf{A}} = \mathbf{U}_\beta^\dagger\mathbf{A}\mathbf{U}_\beta \tag{87}$$

$$\widehat{\boldsymbol{\alpha}} = \mathbf{U}_\beta^\dagger\boldsymbol{\alpha}, \tag{88}$$

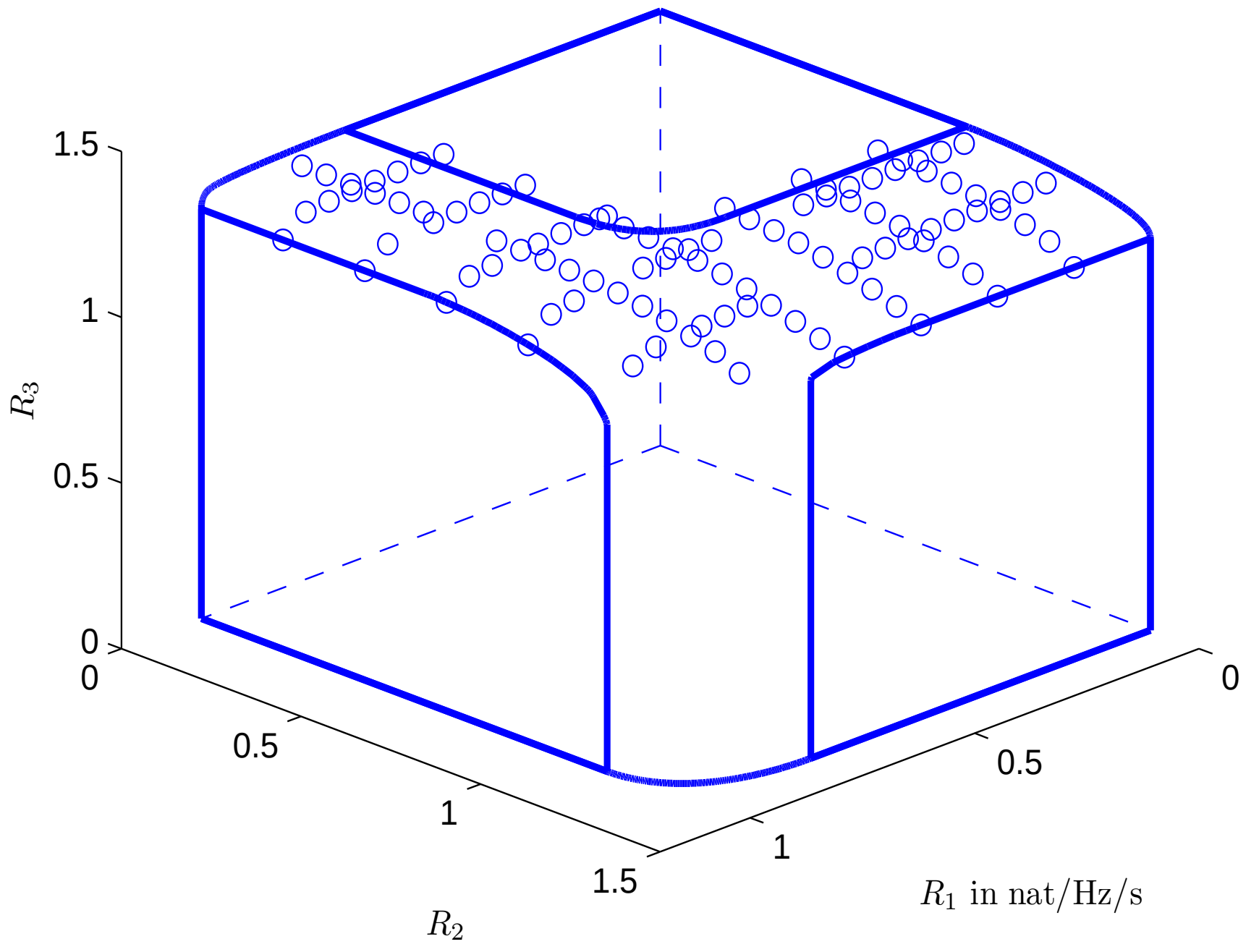

Fig. 6. The SUD rate regions of a three-user MISO IC.

$$\widehat{A}_{11} = \left(\widehat{\mathbf{A}}\right)_{11} \tag{89}$$

$$\text{and} \quad \alpha_1 = (\boldsymbol{\alpha})_1 \,. \tag{90}$$

On substituting (83), (87) and (88) into (81), we have

$$\frac{z_1^2}{\|\boldsymbol{h}_3\|^2}\widehat{\mathbf{A}} \succeq \widehat{\boldsymbol{\alpha}}\widehat{\boldsymbol{\alpha}}^\dagger. \tag{91}$$

We first assume $\widehat{h}_{11} \neq 0$. Then under the constraints in problem (85), we have

$$\begin{aligned}
&\frac{z_1^2}{\|\boldsymbol{h}_3\|^2}\left|\widehat{h}_{11}\right|^2 + \widehat{h}_{11}\boldsymbol{\beta}^\dagger\boldsymbol{\alpha} + \widehat{h}_{11}^\dagger\boldsymbol{\alpha}^\dagger\boldsymbol{\beta} + \boldsymbol{\beta}^\dagger\mathbf{A}\boldsymbol{\beta} \\
&\overset{(a)}{=} \frac{z_1^2}{\|\boldsymbol{h}_3\|^2}\left|\widehat{h}_{11}\right|^2 + \widehat{h}_{11}\begin{bmatrix}\|\boldsymbol{\beta}\| \\ \mathbf{0}\end{bmatrix}^\dagger\widehat{\boldsymbol{\alpha}} + \widehat{h}_{11}^\dagger\widehat{\boldsymbol{\alpha}}^\dagger\begin{bmatrix}\|\boldsymbol{\beta}\| \\ \mathbf{0}\end{bmatrix} + \begin{bmatrix}\|\boldsymbol{\beta}\| \\ \mathbf{0}\end{bmatrix}^\dagger\widehat{\mathbf{A}}\begin{bmatrix}\|\boldsymbol{\beta}\| \\ \mathbf{0}\end{bmatrix} \\
&\overset{(b)}{=} \frac{z_1^2}{\|\boldsymbol{h}_3\|^2}\left|\widehat{h}_{11}\right|^2 + \widehat{h}_{11}\|\boldsymbol{\beta}\|\widehat{\alpha}_1 + \widehat{h}_{11}^\dagger\|\boldsymbol{\beta}\|\widehat{\alpha}_1^\dagger + \|\boldsymbol{\beta}\|^2\widehat{A}_{11} \\
&\overset{(c)}{\leq} \frac{z_1^2}{\|\boldsymbol{h}_3\|^2}\left|\widehat{h}_{11}\right|^2 + \widehat{h}_{11}\|\boldsymbol{\beta}\|\widehat{\alpha}_1 + \widehat{h}_{11}^\dagger\|\boldsymbol{\beta}\|\widehat{\alpha}_1^\dagger + \|\boldsymbol{\beta}\|^2\left(P_1 - \frac{z_1^2}{\|\boldsymbol{h}_3\|^2}\right)
\end{aligned}$$

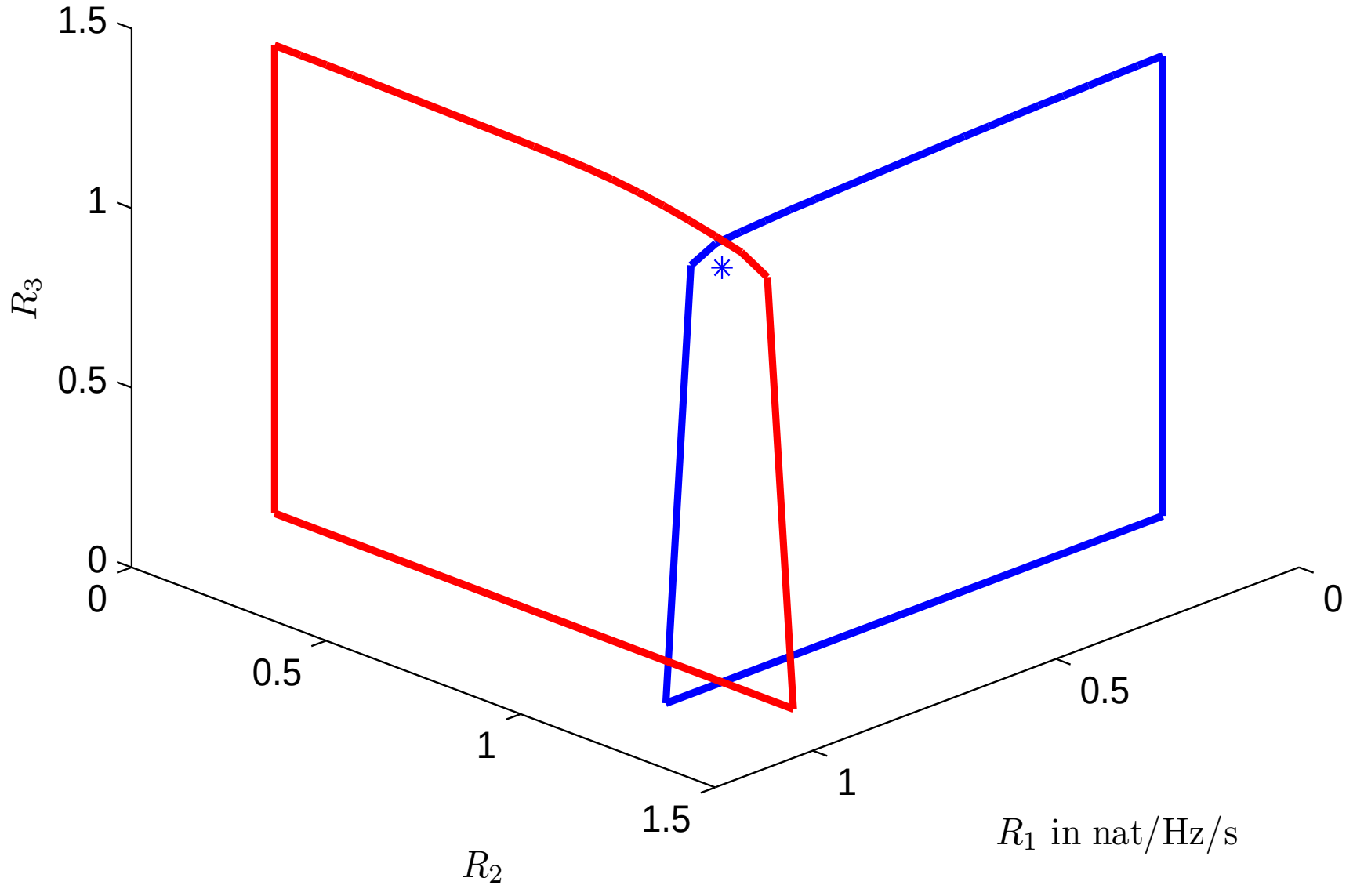

Fig. 7. The ZF beamforming rate triple $R_1 = 1.8118$, $R_2 = 2.2998$ and $R_3 = 2.3077$, and the cutting planes that contain the ZF beamforming rate point and are parallel to $R_3$-$O$-$R_2$ and $R_3$-$O$-$R_1$, respectively.

$$\begin{aligned}
&\overset{(d)}{\leq} \frac{z_1^2}{\|\boldsymbol{h}_3\|^2}\left|\widehat{h}_{11}\right|^2 + 2\|\boldsymbol{\beta}\|\left|\widehat{h}_{11}\right||\widehat{\alpha}_1| + \|\boldsymbol{\beta}\|^2\left(P_1 - \frac{z_1^2}{\|\boldsymbol{h}_3\|^2}\right) \\
&\overset{(e)}{\leq} \frac{z_1^2}{\|\boldsymbol{h}_3\|^2}\left|\widehat{h}_{11}\right|^2 + 2\|\boldsymbol{\beta}\|\left|\widehat{h}_{11}\right|\sqrt{\frac{z_1^2}{\|\boldsymbol{h}_3\|^2}\left(P_1 - \frac{z_1^2}{\|\boldsymbol{h}_3\|^2}\right)} + \|\boldsymbol{\beta}\|^2\left(P_1 - \frac{z_1^2}{\|\boldsymbol{h}_3\|^2}\right) \\
&= \left(\frac{z_1}{\|\boldsymbol{h}_3\|}\left|\widehat{h}_{11}\right| + \|\boldsymbol{\beta}\|\sqrt{P_1 - \frac{z_1^2}{\|\boldsymbol{h}_3\|^2}}\right)^2, \qquad (92)
\end{aligned}$$

where (a) is from (86)-(88); (b) is from (89) and (90); and (c) is from the facts that $\mathbf{A} \succeq \mathbf{0}$ and

$$\widehat{A}_{11} \leq \operatorname{tr}\left(\widehat{\mathbf{A}}\right) = \operatorname{tr}(\mathbf{A}) \leq P_1 - \frac{z_1^2}{\|\boldsymbol{h}_3\|^2}, \qquad (93)$$

and the equality of (c) holds if and only if

$$\widehat{A}_{11} = P_1 - \frac{z_1^2}{\|\boldsymbol{h}_3\|^2}. \qquad (94)$$

Since $\widehat{\mathbf{A}} \succeq \mathbf{0}$, (93) and (94) imply that

$$\widehat{\mathbf{A}} = \begin{bmatrix} P_1 - \frac{z_1^2}{\|\boldsymbol{h}_3\|^2} & \mathbf{0} \\ \mathbf{0} & \mathbf{0} \end{bmatrix}. \tag{95}$$

Inequality (d) is from the fact that

$$\begin{aligned} \widehat{h}_{11}\widehat{\alpha}_1 + \widehat{h}_{11}^{\dagger}\widehat{\alpha}_1^{\dagger} &= \text{Re}\left(\widehat{h}_{11}\widehat{\alpha}_1\right) \\ &\leq 2\left|\widehat{h}_{11}\right| |\widehat{\alpha}_1| . \end{aligned} \tag{96}$$

The equality of (96) as well as the equality of (d) holds if and only if $h_{11}\widehat{\alpha}_1$ is real, i.e.,

$$\widehat{h}_{11}\widehat{\alpha}_1 = \left(\widehat{h}_{11}\widehat{\alpha}_1\right)^{\dagger}. \tag{97}$$

Inequality (e) holds because of (91) and (93):

$$|\widehat{\alpha}_1| \leq \frac{z_1}{\|\boldsymbol{h}_3\|}\sqrt{\widehat{A}_{11}} \leq \frac{z_1}{\|\boldsymbol{h}_3\|}\sqrt{P_1 - \frac{z_1^2}{\|\boldsymbol{h}_3\|^2}}. \tag{98}$$

The equality holds under the condition of (95) and

$$\widehat{\boldsymbol{\alpha}} = \begin{bmatrix} \frac{\widehat{h}_{11}^{\dagger}}{\left|\widehat{h}_{11}\right|}\sqrt{\frac{z_1^2}{\|\boldsymbol{h}_3\|^2}\left(P_1 - \frac{z_1^2}{\|\boldsymbol{h}_3\|^2}\right)} \\ \mathbf{0} \end{bmatrix}, \tag{99}$$

where the choice of $\widehat{\boldsymbol{\alpha}}$ ensures that (97) is satisfied. Thus, in summary, the equality of (92) holds if and only if (95) and (99) are satisfied.

From (87), (88), (95) and (99), the optimal $\boldsymbol{\alpha}$ and $\mathbf{A}$ are

$$\boldsymbol{\alpha} = \mathbf{U}_{\beta}\widehat{\boldsymbol{\alpha}} = \frac{\widehat{h}_{11}^{\dagger}z_1}{\left|\widehat{h}_{11}\right|\|\boldsymbol{h}_3\|}\sqrt{P_1 - \frac{z_1^2}{\|\boldsymbol{h}_3\|^2}} \cdot \frac{\boldsymbol{\beta}}{\|\boldsymbol{\beta}\|} \tag{100}$$

$$\text{and} \quad \mathbf{A} = \frac{\|\boldsymbol{h}_3\|^2}{z_1^2}\boldsymbol{\alpha}\boldsymbol{\alpha}^{\dagger} = \left(P_1 - \frac{z_1^2}{\|\boldsymbol{h}_3\|^2}\right)\frac{\boldsymbol{\beta}\boldsymbol{\beta}^{\dagger}}{\|\boldsymbol{\beta}\|^2}. \tag{101}$$

Therefore, the optimal covariance matrix for problem (4) is

$$\begin{aligned} \mathbf{S}_1^* &= \mathbf{U}_3 \begin{bmatrix} \frac{z_1^2}{\|\boldsymbol{h}_3\|^2} & \frac{\widehat{h}_{11}z_1}{\left|\widehat{h}_{11}\right|\|\boldsymbol{h}_3\|}\sqrt{P_1 - \frac{z_1^2}{\|\boldsymbol{h}_3\|^2}} \cdot \frac{\boldsymbol{\beta}^{\dagger}}{\|\boldsymbol{\beta}\|} \\ \frac{\widehat{h}_{11}^{\dagger}z_1}{\left|\widehat{h}_{11}\right|\|\boldsymbol{h}_3\|}\sqrt{P_1 - \frac{z_1^2}{\|\boldsymbol{h}_3\|^2}} \cdot \frac{\boldsymbol{\beta}}{\|\boldsymbol{\beta}\|} & \left(P_1 - \frac{z_1^2}{\|\boldsymbol{h}_3\|^2}\right)\frac{\boldsymbol{\beta}\boldsymbol{\beta}^{\dagger}}{\|\boldsymbol{\beta}\|^2} \end{bmatrix} \mathbf{U}_3^{\dagger} \\ &= \boldsymbol{\gamma}_1\boldsymbol{\gamma}_1^{\dagger}. \end{aligned} \tag{102}$$

The maximum of problem (4) is

$$\begin{aligned}
&\left(\boldsymbol{h}_1^\dagger \mathbf{S} \boldsymbol{h}_1\right)_{\max} \\
&= \left(\frac{z_1}{\|\boldsymbol{h}_3\|}\left|\widehat{h}_{11}\right| + \|\boldsymbol{\beta}\|\sqrt{P_1 - \frac{z_1^2}{\|\boldsymbol{h}_3\|^2}}\right)^2 \\
&= \left(\frac{z_1\left|\boldsymbol{h}_3^\dagger \boldsymbol{h}_1\right|}{\|\boldsymbol{h}_3\|^2} + \sqrt{\left(\|\boldsymbol{h}_1\|^2 - \frac{1}{\|\boldsymbol{h}_3\|^2}\left|\boldsymbol{h}_3^\dagger \boldsymbol{h}_1\right|^2\right)\left(P_1 - \frac{z_1^2}{\|\boldsymbol{h}_3\|^2}\right)}\right)^2. \qquad (103)
\end{aligned}$$

When $\widehat{h}_{11} = 0$, problem (85) is solved by letting

$$\mathbf{A} = \left(P_1 - \frac{z_1^2}{\|\boldsymbol{h}_3\|^2}\right)\frac{\boldsymbol{\beta}\boldsymbol{\beta}^\dagger}{\|\boldsymbol{\beta}\|^2}.$$

In this case $\boldsymbol{\alpha}$ does not change the value of problem (85). It only needs to satisfy (81). In Lemma 2, we choose $\boldsymbol{\alpha}$ to satisfy the equality of (81) to make the optimal covariance matrix be rank-1. Consequently, the optimal $\mathbf{S}_1$ and the maximum of (4) are respectively (12) and (13) with $k = 1$.

### *B. Proof of Lemma 5*

We first consider the special case of $\|\boldsymbol{y}\| = 0$. Obviously, (35) holds and we can choose $\mathbf{K}^*$ as (38) so that (39) holds. One can also choose $\mathbf{K}_{22} \neq \mathbf{0}$ which still achieves the equality of (35) but violates (39). Therefore, in this case $\mathbf{K}^*$ is not unique unless $\mathrm{tr}(\mathbf{K}_{11}) = P$.

Another special case is $\mathbf{K}_{11} = \mathbf{0}$ and $\|\boldsymbol{y}\| \neq 0$. Since $\mathbf{K} \succeq \mathbf{0}$ we have $\mathbf{K}_{21} = \mathbf{0}$. To achieve the equality of (35), we choose $\mathbf{K}_{22} = P\boldsymbol{y}\boldsymbol{y}^\dagger/\|\boldsymbol{y}\|^2$. Therefore, $\mathrm{rank}\left(\mathbf{K}^*\right) = 1$ and (39) holds.

Next, we need only to prove Lemma 5 when $\mathbf{K}_{11} \neq \mathbf{0}$ and $\|\boldsymbol{y}\| \neq 0$. Let $\mathrm{rank}(\mathbf{K}_{11}) = \mathrm{rank}\left(\boldsymbol{\Lambda}\right) = r$. Then

$$\begin{aligned}
&\begin{bmatrix}\boldsymbol{x}\\ \boldsymbol{y}\end{bmatrix}^\dagger \begin{bmatrix}\mathbf{K}_{11} & \mathbf{K}_{21}^\dagger \\ \mathbf{K}_{21} & \mathbf{K}_{22}\end{bmatrix}\begin{bmatrix}\boldsymbol{x}\\ \boldsymbol{y}\end{bmatrix} \\
&= \begin{bmatrix}\boldsymbol{x}\\ \boldsymbol{y}\end{bmatrix}^\dagger \begin{bmatrix}\mathbf{Q}^\dagger \begin{bmatrix}\boldsymbol{\Lambda} & \mathbf{0}\\ \mathbf{0} & \mathbf{0}\end{bmatrix}\mathbf{Q} & \mathbf{K}_{21}^\dagger \\ \mathbf{K}_{21} & \mathbf{K}_{22}\end{bmatrix}\begin{bmatrix}\boldsymbol{x}\\ \boldsymbol{y}\end{bmatrix} \\
&= \begin{bmatrix}\boldsymbol{x}\\ \boldsymbol{y}\end{bmatrix}^\dagger \begin{bmatrix}\mathbf{Q}^\dagger \begin{bmatrix}\boldsymbol{\Lambda}^{\frac{1}{2}} & \mathbf{0}\\ \mathbf{0} & \mathbf{I}\end{bmatrix} & \mathbf{0} \\ \mathbf{0} & \mathbf{I}\end{bmatrix}
\begin{bmatrix}\begin{bmatrix}\mathbf{I}_{r\times r} & \mathbf{0}\\ \mathbf{0} & \mathbf{0}\end{bmatrix} & \begin{bmatrix}\boldsymbol{\Lambda}^{-\frac{1}{2}} & \mathbf{0}\\ \mathbf{0} & \mathbf{I}\end{bmatrix}\mathbf{Q}\mathbf{K}_{21}^\dagger \\ \mathbf{K}_{21}\mathbf{Q}^\dagger\begin{bmatrix}\boldsymbol{\Lambda}^{-\frac{1}{2}} & \mathbf{0}\\ \mathbf{0} & \mathbf{I}\end{bmatrix} & \mathbf{K}_{22}\end{bmatrix}
\begin{bmatrix}\begin{bmatrix}\boldsymbol{\Lambda}^{\frac{1}{2}} & \mathbf{0}\\ \mathbf{0} & \mathbf{I}\end{bmatrix}\mathbf{Q} & \mathbf{0} \\ \mathbf{0} & \mathbf{I}\end{bmatrix}\begin{bmatrix}\boldsymbol{x}\\ \boldsymbol{y}\end{bmatrix}
\end{aligned}$$

$$\overset{(a)}{=} \begin{bmatrix} \widehat{\boldsymbol{x}}_1 \\ \widehat{\boldsymbol{x}}_2 \\ \boldsymbol{y} \end{bmatrix}^\dagger \begin{bmatrix} \begin{bmatrix} \mathbf{I}_{r\times r} & \mathbf{0} \\ \mathbf{0} & \mathbf{0} \end{bmatrix} & \begin{bmatrix} \boldsymbol{\Lambda}^{-\frac{1}{2}} & \mathbf{0} \\ \mathbf{0} & \mathbf{I} \end{bmatrix} \mathbf{Q}\mathbf{K}_{21}^\dagger \\ \mathbf{K}_{21}\mathbf{Q}^\dagger \begin{bmatrix} \boldsymbol{\Lambda}^{-\frac{1}{2}} & \mathbf{0} \\ \mathbf{0} & \mathbf{I} \end{bmatrix} & \mathbf{K}_{22} \end{bmatrix} \begin{bmatrix} \widehat{\boldsymbol{x}}_1 \\ \widehat{\boldsymbol{x}}_2 \\ \boldsymbol{y} \end{bmatrix} \tag{104}$$

$$\overset{(b)}{=} \begin{bmatrix} \widehat{\boldsymbol{x}}_1 \\ \widehat{\boldsymbol{x}}_2 \\ \boldsymbol{y} \end{bmatrix}^\dagger \begin{bmatrix} \mathbf{I}_{r\times r} & \mathbf{0} & \mathbf{D}^\dagger \\ \mathbf{0} & \mathbf{0} & \mathbf{0} \\ \mathbf{D} & \mathbf{0} & \mathbf{K}_{22} \end{bmatrix} \begin{bmatrix} \widehat{\boldsymbol{x}}_1 \\ \widehat{\boldsymbol{x}}_2 \\ \boldsymbol{y} \end{bmatrix}$$

$$= \begin{bmatrix} \widehat{\boldsymbol{x}}_1 \\ \boldsymbol{y} \end{bmatrix}^\dagger \begin{bmatrix} \mathbf{I}_{r\times r} & \mathbf{D}^\dagger \\ \mathbf{D} & \mathbf{K}_{22} \end{bmatrix} \begin{bmatrix} \widehat{\boldsymbol{x}}_1 \\ \boldsymbol{y} \end{bmatrix}$$

$$\overset{(c)}{=} \begin{bmatrix} \|\widehat{\boldsymbol{x}}_1\| \\ \mathbf{0} \\ \|\boldsymbol{y}\| \\ \mathbf{0} \end{bmatrix}^\dagger \begin{bmatrix} \widehat{\mathbf{U}}_{x_1}^\dagger & \mathbf{0} \\ \mathbf{0} & \mathbf{U}_y^\dagger \end{bmatrix} \begin{bmatrix} \mathbf{I}_{r\times r} & \mathbf{D}^\dagger \\ \mathbf{D} & \mathbf{K}_{22} \end{bmatrix} \begin{bmatrix} \widehat{\mathbf{U}}_{x_1} & \mathbf{0} \\ \mathbf{0} & \mathbf{U}_y \end{bmatrix} \begin{bmatrix} \|\widehat{\boldsymbol{x}}_1\| \\ \mathbf{0} \\ \|\boldsymbol{y}\| \\ \mathbf{0} \end{bmatrix}$$

$$= \begin{bmatrix} \|\widehat{\boldsymbol{x}}_1\| \\ \mathbf{0} \\ \|\boldsymbol{y}\| \\ \mathbf{0} \end{bmatrix}^\dagger \begin{bmatrix} \mathbf{I}_{r\times r} & \widehat{\mathbf{U}}_{x_1}^\dagger \mathbf{D}^\dagger \mathbf{U}_y \\ \mathbf{U}_y^\dagger \mathbf{D} \widehat{\mathbf{U}}_{x_1} & \mathbf{U}_y^\dagger \mathbf{K}_{22} \mathbf{U}_y \end{bmatrix} \begin{bmatrix} \|\widehat{\boldsymbol{x}}_1\| \\ \mathbf{0} \\ \|\boldsymbol{y}\| \\ \mathbf{0} \end{bmatrix}$$

$$= \|\widehat{\boldsymbol{x}}_1\|^2 + 2\|\widehat{\boldsymbol{x}}_1\|\,\|\boldsymbol{y}\| \mathrm{Re}\left[\left(\mathbf{U}_y^\dagger \mathbf{D} \widehat{\mathbf{U}}_{x_1}\right)_{11}\right] + \|\boldsymbol{y}\|^2 \left(\mathbf{U}_y^\dagger \mathbf{K}_{22} \mathbf{U}_y\right)_{11}$$

$$\le \|\widehat{\boldsymbol{x}}_1\|^2 + 2\|\widehat{\boldsymbol{x}}_1\|\,\|\boldsymbol{y}\| \left|\left(\mathbf{U}_y^\dagger \mathbf{D} \widehat{\mathbf{U}}_{x_1}\right)_{11}\right| + \|\boldsymbol{y}\|^2 \left(\mathbf{U}_y^\dagger \mathbf{K}_{22} \mathbf{U}_y\right)_{11}$$

$$\overset{(d)}{\le} \|\widehat{\boldsymbol{x}}_1\|^2 + 2\|\widehat{\boldsymbol{x}}_1\|\,\|\boldsymbol{y}\|\sqrt{P - \mathrm{tr}(\mathbf{K}_{11})} + \|\boldsymbol{y}\|^2 \left(P - \mathrm{tr}(\mathbf{K}_{11})\right),$$

$$\overset{(e)}{=} \left(\sqrt{\boldsymbol{x}^\dagger \mathbf{K}_{11} \boldsymbol{x}} + \|\boldsymbol{y}\|\sqrt{P - \mathrm{tr}(\mathbf{K}_{11})}\right)^2 \tag{105}$$

where in (a) we define

$$\begin{bmatrix} \widehat{\boldsymbol{x}}_1 \\ \widehat{\boldsymbol{x}}_2 \end{bmatrix} = \begin{bmatrix} \boldsymbol{\Lambda}^{\frac{1}{2}} & \mathbf{0} \\ \mathbf{0} & \mathbf{I} \end{bmatrix} \mathbf{Q}\boldsymbol{x}, \tag{106}$$

and $\widehat{\boldsymbol{x}}_1$ is $r \times 1$ and $\widehat{\boldsymbol{x}}_2$ is $(t_1 - r) \times 1$. In (b) we define

$$\begin{bmatrix} \mathbf{D}^\dagger \\ \mathbf{0} \end{bmatrix} = \begin{bmatrix} \boldsymbol{\Lambda}^{-\frac{1}{2}} & \mathbf{0} \\ \mathbf{0} & \mathbf{I} \end{bmatrix} \mathbf{Q}\mathbf{K}_{21}^\dagger. \tag{107}$$

The lower part of the matrix on the right-hand side of (107) must be the all-zero matrix, since the second matrix in the quadratic form (104) is positive semi-definite. In (c), we let the SVD of $\widehat{\boldsymbol{x}}_1$ and $\boldsymbol{y}$ be

respectively

$$\widehat{\boldsymbol{x}}_1 = \widehat{\mathbf{U}}_{x_1} \begin{bmatrix} \|\widehat{\boldsymbol{x}}_1\| \\ \mathbf{0} \end{bmatrix} \tag{108}$$

$$\text{and} \quad \boldsymbol{y} = \mathbf{U}_y \begin{bmatrix} \|\boldsymbol{y}\| \\ \mathbf{0} \end{bmatrix}. \tag{109}$$

Since

$$\begin{bmatrix} \mathbf{I} & \widehat{\mathbf{U}}_{x_1}^{\dagger} \mathbf{D}^{\dagger} \mathbf{U}_y \\ \mathbf{U}_y^{\dagger} \mathbf{D} \widehat{\mathbf{U}}_{x_1} & \mathbf{U}_y^{\dagger} \mathbf{K}_{22} \mathbf{U}_y \end{bmatrix} \succeq \mathbf{0}, \tag{110}$$

we have

$$\left(\mathbf{U}_y^{\dagger} \mathbf{K}_{22} \mathbf{U}_y\right)_{11} \leq \text{tr}\left(\mathbf{U}_y^{\dagger} \mathbf{K}_{22} \mathbf{U}_y\right) = \text{tr}\left(\mathbf{K}_{22}\right) \leq P - \text{tr}\left(\mathbf{K}_{11}\right), \tag{111}$$

and

$$\left|\left(\mathbf{U}_y^{\dagger} \mathbf{D} \widehat{\mathbf{U}}_{x_1}\right)_{11}\right| \leq \sqrt{\left(\mathbf{U}_y^{\dagger} \mathbf{K}_{22} \mathbf{U}_y\right)_{11}} \leq \sqrt{P - \text{tr}\left(\mathbf{K}_{11}\right)}. \tag{112}$$

Therefore (d) holds with equality when

$$\mathbf{U}_y^{\dagger} \mathbf{D} \widehat{\mathbf{U}}_{x_1} = \begin{bmatrix} \sqrt{P - \text{tr}\left(\mathbf{K}_{11}\right)} & \mathbf{0} \\ \mathbf{0} & \mathbf{0}_{(t_2-1)\times(r-1)} \end{bmatrix} \tag{113}$$

$$\text{and} \quad \mathbf{U}_y^{\dagger} \mathbf{K}_{22} \mathbf{U}_y = \begin{bmatrix} P - \text{tr}\left(\mathbf{K}_{11}\right) & \mathbf{0} \\ \mathbf{0} & \mathbf{0}_{(t_2-1)\times(t_2-1)} \end{bmatrix}. \tag{114}$$

From (106) and (108) we have

$$\begin{aligned} \|\widehat{\boldsymbol{x}}_1\|^2 &= \widehat{\boldsymbol{x}}_1^{\dagger} \widehat{\boldsymbol{x}}_1 \\ &= \begin{bmatrix} \widehat{\boldsymbol{x}}_1 \\ \widehat{\boldsymbol{x}}_2 \end{bmatrix}^{\dagger} \begin{bmatrix} \mathbf{I} & \mathbf{0} \\ \mathbf{0} & \mathbf{0} \end{bmatrix} \begin{bmatrix} \widehat{\boldsymbol{x}}_1 \\ \widehat{\boldsymbol{x}}_2 \end{bmatrix} \\ &= \boldsymbol{x}^{\dagger} \mathbf{Q}^{\dagger} \begin{bmatrix} \mathbf{\Lambda}^{\frac{1}{2}} & \mathbf{0} \\ \mathbf{0} & \mathbf{0} \end{bmatrix} \begin{bmatrix} \mathbf{I} & \mathbf{0} \\ \mathbf{0} & \mathbf{0} \end{bmatrix} \begin{bmatrix} \mathbf{\Lambda}^{\frac{1}{2}} & \mathbf{0} \\ \mathbf{0} & \mathbf{0} \end{bmatrix} \mathbf{Q} \boldsymbol{x} \\ &= \boldsymbol{x}^{\dagger} \mathbf{Q}^{\dagger} \begin{bmatrix} \mathbf{\Lambda} & \mathbf{0} \\ \mathbf{0} & \mathbf{0} \end{bmatrix} \mathbf{Q} \boldsymbol{x} \\ &= \boldsymbol{x}^{\dagger} \mathbf{K}_{11} \boldsymbol{x}. \end{aligned} \tag{115}$$

Therefore, (e) holds. Thus, in summary, the equality of (105) holds when (113) and (114) hold. Therefore, (34) is true and the optimal $\mathbf{K}^*$ satisfies

$$\begin{aligned}\text{rank}\left(\mathbf{K}^*\right) &= \text{rank}\begin{bmatrix}\mathbf{I} & \widehat{\mathbf{U}}_{x_1}^\dagger \mathbf{D}^\dagger \mathbf{U}_y \\ \mathbf{U}_y^\dagger \mathbf{D}\widehat{\mathbf{U}}_{x_1} & \mathbf{U}_y^\dagger \mathbf{K}_{22}\mathbf{U}_y\end{bmatrix} \\ &= \text{rank}\begin{bmatrix}\multicolumn{2}{c}{\mathbf{I}_{r\times r}} & \sqrt{P-\text{tr}\left(\mathbf{K}_{11}\right)} & \mathbf{0} \\ & & \mathbf{0} & \mathbf{0} \\ \sqrt{P-\text{tr}\left(\mathbf{K}_{11}\right)} & \mathbf{0} & P-\text{tr}\left(\mathbf{K}_{11}\right) & \mathbf{0} \\ \mathbf{0} & \mathbf{0} & \mathbf{0} & \mathbf{0}\end{bmatrix} \\ &= r \\ &= \text{rank}\left(\mathbf{K}_{11}\right).\end{aligned} \tag{116}$$

In the following we obtain $\mathbf{K}$ that achieves the equality of (35).

1) When $\boldsymbol{x}^\dagger\mathbf{K}_{11}\boldsymbol{x} \neq 0$, from (113) and (114) we have

$$\mathbf{D} = \mathbf{U}_y \begin{bmatrix}\sqrt{P-\text{tr}\left(\mathbf{K}_{11}\right)} & \mathbf{0} \\ \mathbf{0} & \mathbf{0}\end{bmatrix}\widehat{\mathbf{U}}_{x_1}^\dagger = \frac{\sqrt{P-\text{tr}\left(\mathbf{K}_{11}\right)}}{\|\widehat{\boldsymbol{x}}_1\|\cdot\|\boldsymbol{y}\|}\boldsymbol{y}\widehat{\boldsymbol{x}}_1^\dagger \tag{117}$$

$$\text{and} \quad \mathbf{K}_{22} = \mathbf{U}_y \begin{bmatrix}P-\text{tr}\left(\mathbf{K}_{11}\right) & \mathbf{0} \\ \mathbf{0} & \mathbf{0}\end{bmatrix}\mathbf{U}_y^\dagger = \frac{P-\text{tr}\left(\mathbf{K}_{11}\right)}{\|\boldsymbol{y}\|^2}\boldsymbol{y}\boldsymbol{y}^\dagger. \tag{118}$$

From (107) and (117), we have

$$\begin{aligned}\mathbf{K}_{21} &= [\mathbf{D}\quad \mathbf{0}]\begin{bmatrix}\mathbf{\Lambda}^{\frac{1}{2}} & \mathbf{0} \\ \mathbf{0} & \mathbf{I}\end{bmatrix}\mathbf{Q} \\ &= \begin{bmatrix}\frac{\sqrt{P-\text{tr}\left(\mathbf{K}_{11}\right)}}{\|\widehat{\boldsymbol{x}}_1\|\cdot\|\boldsymbol{y}\|}\boldsymbol{y}\widehat{\boldsymbol{x}}_1^\dagger & \mathbf{0}\end{bmatrix}\begin{bmatrix}\mathbf{\Lambda}^{\frac{1}{2}} & \mathbf{0} \\ \mathbf{0} & \mathbf{I}\end{bmatrix}\mathbf{Q} \\ &= \frac{\sqrt{P-\text{tr}\left(\mathbf{K}_{11}\right)}}{\|\widehat{\boldsymbol{x}}_1\|\,\|\boldsymbol{y}\|}\boldsymbol{y}\begin{bmatrix}\widehat{\boldsymbol{x}}_1 \\ \mathbf{0}\end{bmatrix}^\dagger\begin{bmatrix}\mathbf{\Lambda}^{\frac{1}{2}} & \mathbf{0} \\ \mathbf{0} & \mathbf{I}\end{bmatrix}\mathbf{Q} \\ &= \frac{\sqrt{P-\text{tr}\left(\mathbf{K}_{11}\right)}}{\|\widehat{\boldsymbol{x}}_1\|\,\|\boldsymbol{y}\|}\boldsymbol{y}\begin{bmatrix}\widehat{\boldsymbol{x}}_1 \\ \widehat{\boldsymbol{x}}_2\end{bmatrix}^\dagger\begin{bmatrix}\mathbf{I} & \mathbf{0} \\ \mathbf{0} & \mathbf{0}\end{bmatrix}\begin{bmatrix}\mathbf{\Lambda}^{\frac{1}{2}} & \mathbf{0} \\ \mathbf{0} & \mathbf{I}\end{bmatrix}\mathbf{Q} \\ &= \frac{\sqrt{P-\text{tr}\left(\mathbf{K}_{11}\right)}}{\|\widehat{\boldsymbol{x}}_1\|\,\|\boldsymbol{y}\|}\boldsymbol{y}\boldsymbol{x}^\dagger\mathbf{Q}^\dagger\begin{bmatrix}\mathbf{\Lambda}^{\frac{1}{2}} & \mathbf{0} \\ \mathbf{0} & \mathbf{I}\end{bmatrix}\begin{bmatrix}\mathbf{I} & \mathbf{0} \\ \mathbf{0} & \mathbf{0}\end{bmatrix}\begin{bmatrix}\mathbf{\Lambda}^{\frac{1}{2}} & \mathbf{0} \\ \mathbf{0} & \mathbf{I}\end{bmatrix}\mathbf{Q} \\ &= \frac{\sqrt{P-\text{tr}\left(\mathbf{K}_{11}\right)}}{\|\widehat{\boldsymbol{x}}_1\|\,\|\boldsymbol{y}\|}\boldsymbol{y}\boldsymbol{x}^\dagger\mathbf{Q}^\dagger\begin{bmatrix}\mathbf{\Lambda} & \mathbf{0} \\ \mathbf{0} & \mathbf{0}\end{bmatrix}\mathbf{Q}\end{aligned}$$

$$= \frac{\sqrt{P - \text{tr}(\mathbf{K}_{11})}}{\|\widehat{\boldsymbol{x}}_1\| \, \|\boldsymbol{y}\|} \boldsymbol{y}\boldsymbol{x}^\dagger \mathbf{K}_{11}$$
$$= \frac{\sqrt{P - \text{tr}(\mathbf{K}_{11})}}{\|\boldsymbol{y}\| \sqrt{\boldsymbol{x}^\dagger \mathbf{K}_{11} \boldsymbol{x}}} \boldsymbol{y}\boldsymbol{x}^\dagger \mathbf{K}_{11}, \tag{119}$$

where the last equality is from (115).

2) When $\boldsymbol{x}^\dagger \mathbf{K}_{11} \boldsymbol{x} = 0$, $\mathbf{K}_{22}$ is still (118). However, $\widehat{\mathbf{U}}_{x_1}$ can be any unitary matrix since $\|\widehat{\boldsymbol{x}}_1\| = 0$. Therefore, there are different choices of $\mathbf{K}_{21}$ that achieve the equality of (35). We choose $\widehat{\mathbf{U}}_{x_1} = \mathbf{I}$ for convenience. Then, from (107) and (117), we have

$$\mathbf{K}_{21} = \frac{\sqrt{P - \text{tr}(\mathbf{K}_{11})}}{\|\boldsymbol{y}\|} \boldsymbol{y}\mathbf{1}_0^T \begin{bmatrix} \mathbf{\Lambda}^{\frac{1}{2}} & \mathbf{0} \\ \mathbf{0} & \mathbf{I} \end{bmatrix} \mathbf{Q}$$
$$= \frac{\sqrt{P - \text{tr}(\mathbf{K}_{11})}}{\|\boldsymbol{y}\|} \boldsymbol{y}\mathbf{1}_0^T \begin{bmatrix} \mathbf{I}_{r\times r} & \mathbf{0} \\ \mathbf{0} & \mathbf{0} \end{bmatrix} \begin{bmatrix} \mathbf{\Lambda}^{\frac{1}{2}} & \mathbf{0} \\ \mathbf{0} & \mathbf{I} \end{bmatrix} \mathbf{Q}$$
$$= \frac{\sqrt{P - \text{tr}(\mathbf{K}_{11})}}{\|\boldsymbol{y}\|} \boldsymbol{y}\mathbf{1}_0^T \mathbf{K}_{11}^{\frac{1}{2}}. \tag{120}$$